\documentclass[aps,showpacs,preprintnumbers,nofootinbib]{revtex4}%
\usepackage{amssymb}
\usepackage{amsmath}
\usepackage{bm}
\usepackage{amsfonts}
\usepackage{graphicx}
\usepackage{subfig}
\usepackage{placeins}%
\providecommand{\U}[1]{\protect\rule{.1in}{.1in}}
\graphicspath{{figures/}}
\begin{document}
\title{Cosmological models in Weyl geometrical scalar-tensor theory}
\author{M. L. Pucheu, F. A. P. Alves Junior, A. B. Barreto, C. Romero \ \ }
\affiliation{Departamento de F\'{\i}sica, Universidade Federal da Para\'{\i}ba, Caixa
Postal 5008, 58059-970 Jo\~{a}o Pessoa, PB, Brazil }
\affiliation{E-mail: cromero@fisica.ufpb.br }

\begin{abstract}
We investigate cosmological models in a recently proposed geometrical theory
of gravity, in which the scalar field appears as part of the space-time
geometry. We extend the previous theory to include a scalar potential in the
action. We solve the vacuum field equations for different choices of the
scalar potential and give a detailed analysis of the solutions. We show that
in some cases a cosmological scenario is found that seems to suggest the
appearance of a geometric phase transition. We build a toy model, in which the
accelerated expansion of the early universe is driven by pure geometry.

\end{abstract}

\maketitle

\vskip .5cm

\section{Introduction}

Scalar-tensor theories have attracted the attention of cosmologists since a
seminal work by Jordan was published in the fifties \cite{Jordan}. However,
the great impetus to investigate these kinds of theories came from Brans-Dicke
theory, which is considered by many as the most popular and simplest
alternative theory of gravity \cite{Brans}. In the last three decades,
interest in scalar-tensor theory has increased, mainly motivated by modern
Kaluza-Klein theory, string theory, inflationary models, and other recent
proposals. As is well known, scalar-tensor theories in general do not assign
an intrinsic geometric character to the scalar field. Nor does this field
describe matter in the usual sense. In the particular case of Brans-Dicke
theory, its role is to account for possible variations of Newton's
gravitational \textit{constant}, since the latter, according to Mach ideas,
would depend on the mass of the universe \cite{Faraoni}. However, there have
been some attempts to construct a scalar-tensor theory of gravitation in which
the scalar field is an essential part of the space-time geometry
\cite{Peters}. In all these cases, one looks for a geometrical framework that
contains a scalar field as a way of adding a new degree of freedom to the
corresponding gravity theory.

In the present article, we consider a scalar-tensor theory of gravity, in
which the scalar field does play a geometric role. The mechanism for inserting
the scalar field in the space-time geometry is inspired by Weyl's
generalization of Riemannian geometry, in the very special case when the gauge
field corresponds to an exact 1-form \cite{Weyl}. The space-time structure
that results from this kind of geometry is known in the literature as a Weyl
integrable space-time (WIST) \cite{Weyl geometry}. The first approach to a
scalar-tensor theory set in a WIST was proposed by Novello et al, and consists
of a direct extension of general relativity by including in the
Einstein-Hilbert action a term corresponding to a massless scalar field, the
latter being interpreted as a geometrical field in the sense of Weyl geometry
\cite{Novello}. A more recent geometrical approach to scalar-tensor theory
(GST) starts by considering the action of Brans-Dicke theory and introduces
the space-time geometry from first principles, that is, by applying the
Palatini formalism, which then leads to a Weyl integrable geometry
\cite{Pucheu}. Although the original version of GST theory does not include a
scalar potential in the action, the inclusion of such term is rather natural
and does not alter the geometrical aspects of the theory. Thus, in this paper
we consider a slightly modified version of original proposal by adding a
scalar potential. Certainly, the main motivation for this modification lies in
the fact that in modern cosmology scalar potentials are an important
ingredient of inflationary models, quintessence, and other theories.

This article is organized as follows. In Section 2, we give a brief review of
the geometrical scalar-tensor theory and consider its extension to include a
scalar potential. In Section 3, we discuss the application of the geometric
scalar-tensor theory to some cosmological scenarios, in which matter can be
neglected as the potential energy of the scalar field is the dominant
contribution. In Section 4, we construct a very simple cosmological model, a
sort of \textquotedblleft toy model\textquotedblright, in which a phase of
accelerated expansion of the universe is driven by pure geometry. We conclude
in Section V with some remarks.

\section{The Weyl geometrical scalar-tensor theory \qquad}

The Weyl geometrical scalar-tensor theory starts with the action given by
\begin{equation}
S=\int d^{4}x\sqrt{-g}\left\{  e^{-\phi}\left[  R+\omega(\phi)\phi^{,\alpha
}\phi_{,\alpha}\right]  -V(\phi)\right\}  +S_{m}(g,\psi), \label{action1}%
\end{equation}
where $R=g^{\mu\nu}R_{\mu\nu}(\Gamma)$, $\phi$ is a scalar field, $\omega$ is
a function of $\phi$, $V(\phi)$ represents the scalar field potential, and
$S_{m}$ indicates the part of the action depending on the matter fields, here
generically denoted by $\psi$ \footnote{Note that this action is a simple
extension of the action considered in \cite{Pucheu}}. Let us recall that
$\phi$ is regarded as a purely geometrical field, whose meaning becomes clear
only after a Palatini variation of the action above is carried out. Indeed, it
is known that the variation of (\ref{action1}) with respect to the affine
connection $\Gamma_{\mu\nu}^{\alpha}$ leads to \cite{Pucheu}
\begin{equation}
\nabla_{\alpha}g_{\mu\nu}=g_{\mu\nu}\phi_{,\alpha}\text{ ,}
\label{nonmetricity}%
\end{equation}
an equation that expresses the so-called Weyl compatibility condition between
the metric and the connection (also called Weyl nonmetricity
condition)\footnote{Throughout the paper we shall use the following
convention: Whenever the symbol $g$ appears in the expression $\sqrt{-g}$ it
denotes $\det g.$ Otherwise $g$ denotes the metric tensor. \ We shall also
consider the Ricci tensor $R_{\mu\nu}(\Gamma)$\ \ as being given in terms of
the affine connection coefficients $\Gamma_{\mu\nu}^{\alpha}$ via the
definition of the curvature tensor.}. This is the geometric condition that
characterizes the space-time manifold as a Weyl integrable space-time
\cite{Weyl geometry}. By performing the Palatini variation with respect to the
metric $g_{\mu\nu}$ and the scalar field $\phi$ we obtain the following set of
field equations:
\begin{equation}
G_{\mu\nu}=\omega(\phi)\left(  \frac{\phi_{,\alpha}\phi^{,\alpha}}{2}g_{\mu
\nu}-\phi_{,\mu}\phi_{,\nu}\right)  -\frac{1}{2}e^{\phi}g_{\mu\nu}%
V(\phi)-\kappa T_{\mu\nu}, \label{poster06}%
\end{equation}

\begin{equation}
\square\phi=-\left(  1+\frac{1}{2\omega}\frac{d\omega}{d\phi}\right)
\phi_{,\mu}\phi^{,\mu}-\frac{e^{\phi}}{\omega}\left(  \frac{1}{2}\frac
{dV}{d\phi}+V\right)  , \label{poster07}%
\end{equation}
where the symbol $\square$ denotes the d'Alembertian operator calculated with
respect to the Weyl connection, $\kappa=\frac{8\pi}{c^{4}}$, and $T_{\mu\nu}$
represents the Weyl invariant energy-momentum tensor of the matter fields as
defined in \cite{Pucheu}.

\subsection{The field equations in the Riemann frame}

As is well known, the Weyl condition (\ref{nonmetricity}) does not change when
we perform the following transformations in $g$ and $\phi$:%
\begin{equation}
\overline{g}=e^{f}g, \label{conformal}%
\end{equation}%
\begin{equation}
\overline{\phi}=\phi+f. \label{scalar}%
\end{equation}
where $f$ is an arbitrary scalar function defined on the manifold space-time
$M$. These\ transformations are known, in the literature, as Weyl
transformations. The set $(M,g,\phi)$ consisting of a differentiable manifold
$M$ endowed with a metric $g$ and a Weyl scalar field $\phi$ $\ $will be
called a \textit{Weyl frame}. We now note that if we set $f=-\phi$ in
(\ref{scalar}), we get $\overline{\phi}=0$. In this case, when the Weyl scalar
field vanishes, the set $(M,\gamma=e^{-\phi}g,\overline{\phi}=0)$ \ is
referred to as the \textit{Riemann frame}.

It is sometimes convenient to recast the action (\ref{action1}) and the above
field equations in the Riemann frame. It is not difficult to verify that in
this frame (\ref{action1}) is transformed into the action
\begin{equation}
\overline{S}=\int d^{4}x\sqrt{-\gamma}\left[  \bar{R}+\omega(\phi)\gamma
^{\mu\nu}\phi_{,\mu}\phi_{,\nu}-e^{2\phi}V(\phi)\right]  +S^{(m)}(\gamma
,\psi), \label{action}%
\end{equation}
whereas the field equations (\ref{poster06}) and (\ref{poster07}) are given,
respectively, by%

\begin{equation}
\bar{G}_{\mu\nu}=\omega(\phi)\left(  \frac{\phi_{,\alpha}\phi^{,\alpha}}%
{2}\gamma_{\mu\nu}-\phi_{,\mu}\phi_{,\nu}\right)  -\frac{e^{2\phi}}{2}%
\gamma_{\mu\nu}V(\phi)-\kappa T_{\mu\nu}(\gamma), \label{poster09}%
\end{equation}%
\begin{equation}
\bar{\square}\phi=-\frac{1}{2\omega}\frac{d\omega}{d\phi}\phi_{,\alpha}%
\phi^{,\alpha}-\frac{e^{2\phi}}{\omega}\left(  V+\frac{1}{2}\frac{dV}{d\phi
}\right)  , \label{poster10}%
\end{equation}
where both the Einstein tensor $\bar{G}_{\mu\nu}$ and the operator
$\bar{\square}$ are calculated with the Levi-Civita connection given in terms
of the metric $\gamma_{\mu\nu}$.

\section{Applications to cosmology}

Typical gravitational problems, such as the field generated by a spherically
symmetric matter distribution, or the existence of naked singularities and
wormholes as geometric phenomena, have already been studied in the context of
Weyl geometrical scalar-tensor theory \cite{Pucheu}. In this work, we would
like to consider some cosmological models arising from different choices of
the scalar potential. For convenience, we shall work in the Riemann frame,
although due to frame invariance all physical results obtained will be valid
in any Weyl frame \cite{Pucheu}. Let us point out that the technique of frame
transformations to investigate cosmological models with non-minimally coupled
scalar fields is not new, and has been used recently \cite{Kamen}.

As we have already mentioned, scalar fields have been extensively used in
cosmology, mainly motivated by inflationary models, but also as a possible way
to account for dark matter and quintessence models of dark energy
\cite{Matos,Paliathanasis}. In inflationary cosmology, the scalar field
(\textit{the inflaton}) is responsible for the negative pressure needed to
expand the universe \cite{Guth}. Nevertheless, up to now the nature of this
scalar field is not known. On the other hand, since the apperance of
inflationary cosmology in the 1980's different models have been proposed, in
which the presence of a scalar potential $V(\phi)$ is considered \cite{Linde}.
The simplest of these requires a monomial potential. Despite the fact that the
latest observational results do not favour a potential of the type
$V(\phi)\propto\phi^{2}$ \cite{Planck}, a massive scalar field has been
considered by many authors with great interest \cite{piran}. In the following
sections, we shall consider some simple models for different types of
$V(\phi)$ in the context of the Weyl geometrical scalar-tensor theory. In
almost cases we shall add the cosmological constant $\Lambda$. Throughout our
discussion, we shall always take the point of view that the scalar field has
an essentially geometric origin. This is because it is always possible to
interpret the scalar field as a geometric field by going to the Weyl frame
\cite{Pucheu}.

Let us now start by considering a homogeneous, spatially flat, and isotropic
model, whose line element is written as
\begin{equation}
ds^{2}=dt^{2}-a^{2}(t)\left(  dr^{2}+r^{2}d\theta^{2}+r^{2}\sin^{2}\theta
d\phi^{2}\right)  ,\label{poster11}%
\end{equation}
where $a(t)$ denotes the scale factor. Let us also restrict ourselves to the
vacuum case and, as in Brans-Dicke theory, let us for simplicity set
$\omega(\phi)=\omega=const$. Thus, the field equations (\ref{poster09}) reduce
to
\begin{equation}
3\frac{\dot{a}^{2}}{a^{2}}=\frac{\omega}{2}\dot{\phi}^{2}+\frac{e^{2\phi}}%
{2}V(\phi),\label{poster12}%
\end{equation}%
\begin{equation}
2\frac{\ddot{a}}{a}+\frac{\dot{a}^{2}}{a^{2}}=-\frac{\omega}{2}\dot{\phi}%
^{2}+\frac{e^{2\phi}}{2}V(\phi),\label{poster13}%
\end{equation}
while (\ref{poster10}) gives
\begin{equation}
\ddot{\phi}+3\frac{\dot{a}}{a}\dot{\phi}=-\frac{e^{2\phi}}{\omega}\left(
V(\phi)+\frac{1}{2}\frac{dV}{d\phi}\right)  .\label{poster14}%
\end{equation}
Expressing the above equations in terms of the expansion parameter
$\dot{\theta}=\frac{3\dot{a}(t)}{a(t)}$, and defining $\psi=\dot{\phi}$ , it
is easy to verify, after simple calculations, that we are left with the
following equations \footnote{Because the connection $\nabla$ is
frame-invariant, then the expansion parameter $\theta$ is also invariant.
Indeed, by definition $\theta=\nabla_{\mu}U^{\mu}$, where $U$ denotes the
4-velocity field of the fundamental observers. Clearly, in the Riemannian
frame the metric takes the form $\gamma=e^{-\phi}g$, which is, as we know,
invariant. On the other hand, $\gamma$ is assumed to be given by
(\ref{poster11}).}:
\begin{equation}
\frac{\theta^{2}}{3}=\frac{\omega}{2}\psi^{2}+\frac{e^{2\phi}}{2}%
V(\phi),\label{eq01}%
\end{equation}%
\begin{equation}
\dot{\theta}=-\frac{\theta^{2}}{2}-\frac{3\omega}{4}\psi^{2}+\frac{3e^{2\phi}%
}{4}V(\phi),\label{poster31}%
\end{equation}%
\begin{equation}
\dot{\psi}=-\theta\psi-\frac{e^{2\phi}}{\omega}\left(  V(\phi)+\frac{1}%
{2}\frac{dV}{d\phi}\right)  \label{eq3}%
\end{equation}
In the next subsections we shall consider four distinct cases, each one
corresponding to a specific choice of the scalar potential $V(\phi)$.

\subsection{The Cosmological Constant}

We shall start our analysis with the choice $V(\phi)=\Lambda e^{-2\phi}$ .
Clearly, this will lead us, in the Riemann frame, to the case of a massless
scalar field minimally coupled to gravity in the presence of a cosmological
constant $\Lambda$ and a free parameter $\omega$. It is not difficult to check
that, for $\Lambda>0$, the field equations (\ref{poster12}), (\ref{poster13})
and (\ref{poster14}) admit the following solutions:
\begin{equation}
a_{\pm}(t)=a_{0}\exp\left(  \pm\sqrt{\frac{\Lambda}{6}}(t-t_{0})\right)
,\quad\phi=\phi_{0}=const, \label{cosmological_const01}%
\end{equation}
where $a_{0}$ is a constant of integration. These are, in fact, the simplest
of all solutions of the above field equations. The solution $a_{\pm}(t)$
corresponds to the well-known de Sitter (anti-de Sitter) universe, a maximally
symmetric vacuum solution of Einstein's field equations with a cosmological constant.

For $\omega<0$ and $\Lambda>0$, we have two solutions:
\begin{equation}
a(t)=a_{0}\cosh\left(  \sqrt{\frac{3\Lambda}{2}}(t-t_{0})\right)  ^{1/3},
\label{cosmological_const02}%
\end{equation}%
\begin{equation}
\phi(t)=\phi_{0}\pm\sqrt{-\frac{2}{3\omega}}\arctan\left[  \sinh\left(
\sqrt{\frac{3\Lambda}{2}}(t-t_{0})\right)  \right]  .
\label{cosmological_const03}%
\end{equation}
These represent a non-singular bouncing universe which bears some similarity
to recently proposed models in scalar-tensor theories \cite{Stein}. On the
other hand, if $\omega>0$ and $\Lambda<0$ the solutions are
\begin{equation}
a(t)=a_{0}\bigg|\cos\left(  \sqrt{-\frac{3\Lambda}{2}}(t-t_{0})\right)
\bigg|^{1/3}, \label{cosmological_const04}%
\end{equation}%
\begin{equation}
\phi(t)=\phi_{0}\pm\sqrt{\frac{2}{3\omega}}\ln\bigg|\sec\left(  \sqrt
{-\frac{3\Lambda}{2}}(t-t_{0})\right)  +\tan\left(  \sqrt{-\frac{3\Lambda}{2}%
}(t-t_{0})\right)  \bigg|. \label{cosmological_const05}%
\end{equation}
In this case, we have a model that describes a cyclic universe, which
undergoes an eternal series of oscillations, each beginning with a big bang
and ending with a big crunch, and in each cycle a period of expansion is
followed by a contraction. Let us just remark here that cyclic universes have
been favoured by many recent proposals \cite{Turok}. In particular, cyclic
models have also been predicted by loop quantum cosmology through a mechanism
by which the contracting and expanding cosmological branches are connected by
a \textquotedblleft quantum bridge\textquotedblright\ \cite{Asthekar}.

Finally, if both $\Lambda$ and $\omega$ are positive, the solutions will be
given by
\begin{equation}
a(t)=a_{0}\sinh\bigg|\sqrt{\frac{3\Lambda}{2}}(t-t_{0})\bigg|^{1/3},
\label{cosmological_const06}%
\end{equation}%
\begin{equation}
\phi(t)=\phi_{0}\pm\sqrt{\frac{2}{3\omega}}\ln\bigg|\tanh\left(  \frac{1}%
{2}\sqrt{\frac{3\Lambda}{2}}(t-t_{0})\right)  \bigg|.
\label{cosmological_const07}%
\end{equation}
These solutions are singular at $t=t_{0}$ and describe an expanding and
accelerating universe which tends to a de Sitter universe when $t\rightarrow
\infty.$

A nice picture of the time evolution of the above models is given by the phase
portraits of the dynamical system corresponding to the field equations, since
these portraits may provide an insight on the dynamical behaviour of the
solutions. In the next section, we shall give a qualitative analysis of
the solutions obtained.

\subsection{Phase portrait of the solutions when $V(\phi)=\Lambda e^{-2\phi}$}

In this case, it is easy to verify that the field equations (\ref{poster12})-(\ref{poster13}) and (\ref{poster14}) reduce to
\begin{equation}
\frac{\theta^{2}}{3}=\frac{\omega}{2}\psi^{2}+\frac{\Lambda}{2},
\label{poster15}%
\end{equation}%
\begin{equation}
\dot{\theta}=-\frac{1}{2}\theta^{2}-\frac{3\omega}{4}\psi^{2}+\frac{3\Lambda
}{4}, \label{poster16}%
\end{equation}%
\begin{equation}
\dot{\psi}=-\theta\psi. \label{poster17}%
\end{equation}
As we can see, the above equations constitute an autonomous planar dynamical
system, with (\ref{poster15}) representing an algebraic constraint in the
phase space \cite{Andronov}. \FloatBarrier
\begin{figure}[tbh]
\centering
\subfloat[$\omega<0$ and $\Lambda>0$]{
\includegraphics[scale=0.7]{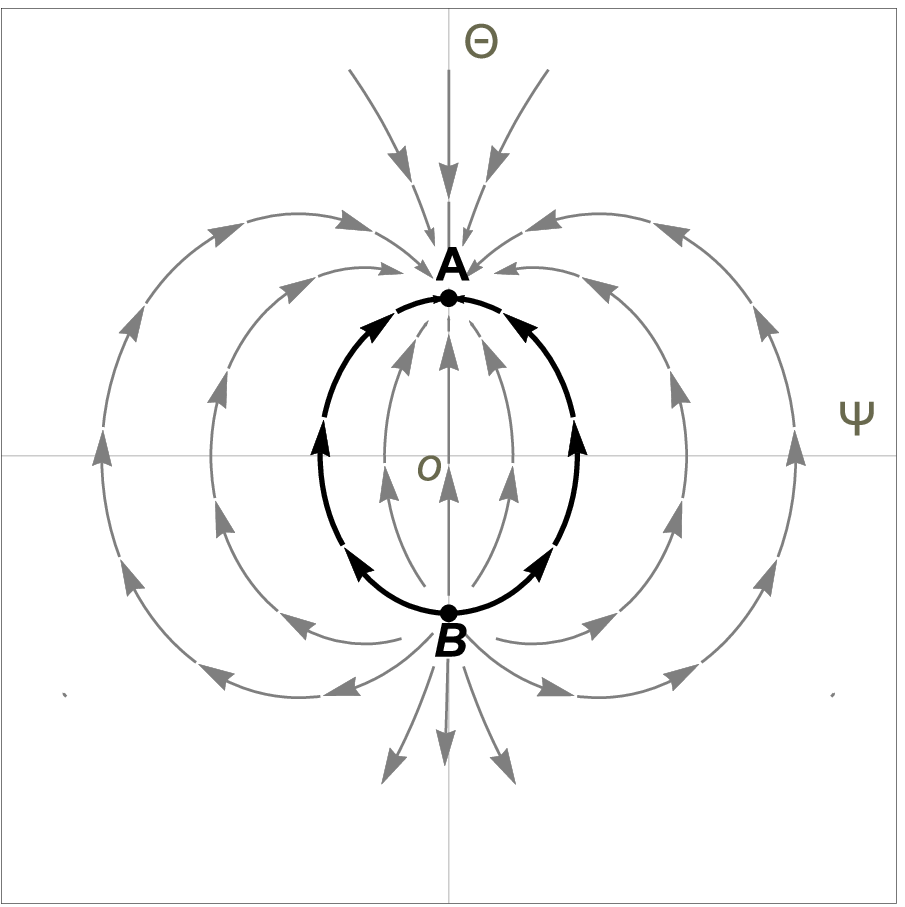}
} \quad
\subfloat[$\omega>0$ and $\Lambda<0$]{
\includegraphics[scale=0.7]{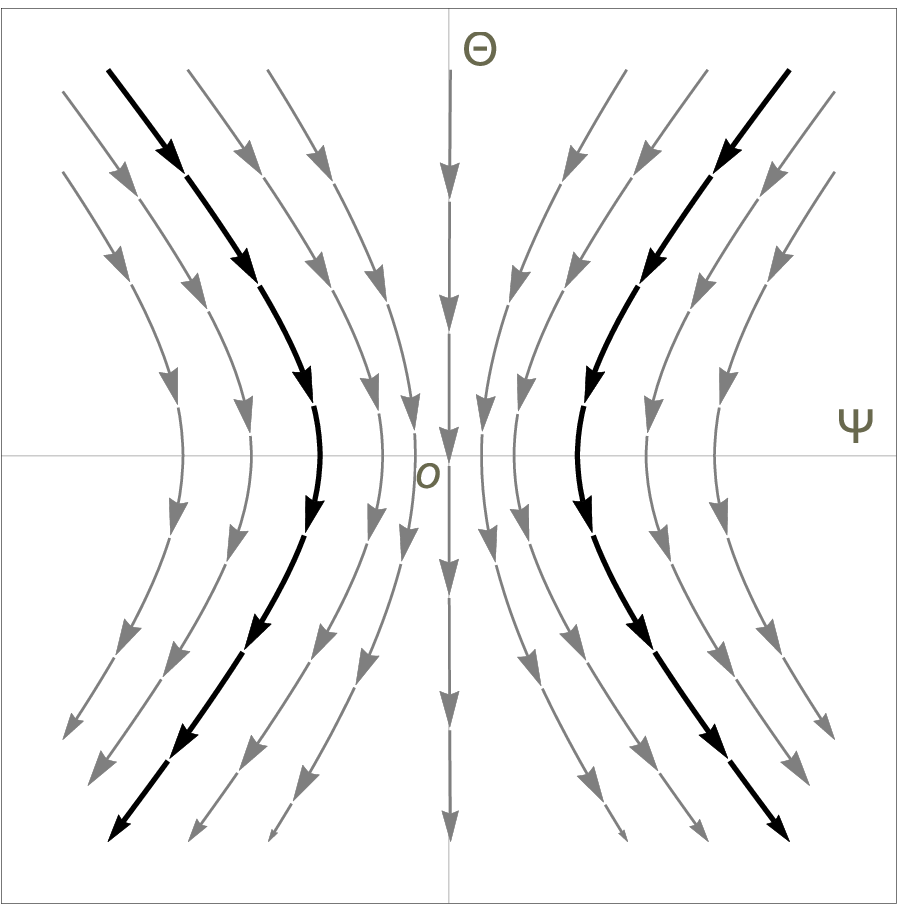}} \quad
\subfloat[$\omega>0$ and $\Lambda>0$]{
\includegraphics[scale=0.7]{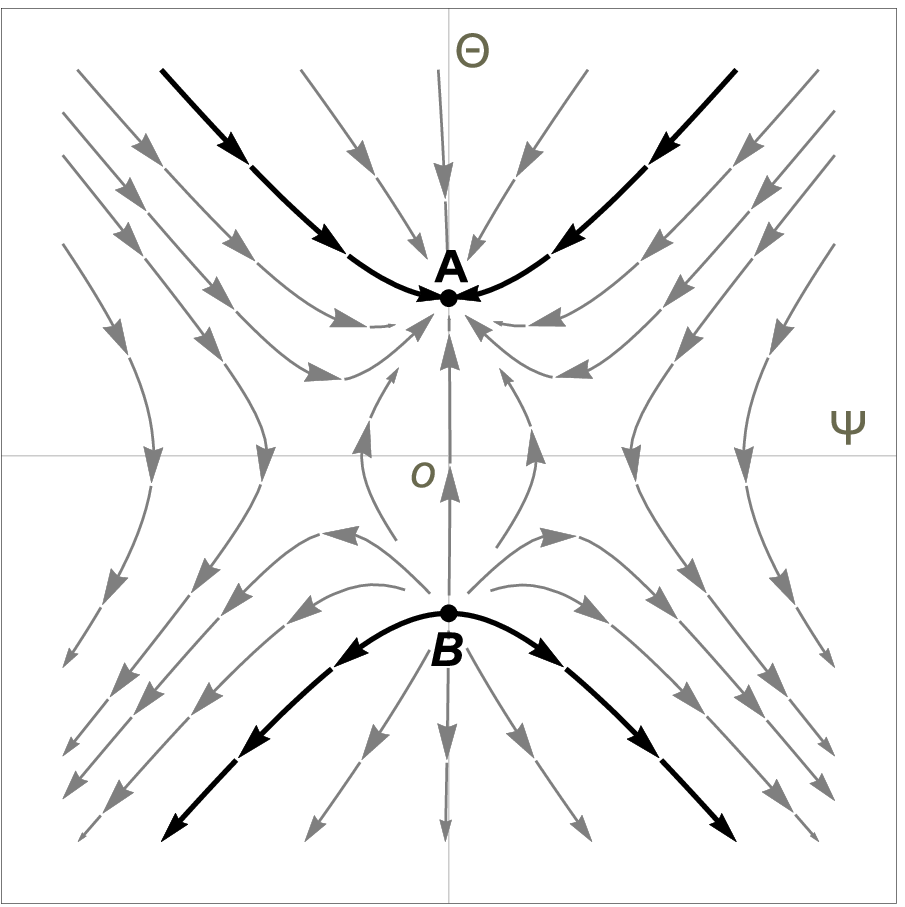}
} \quad\subfloat[$\omega>0$ and $\Lambda=0$]{
\includegraphics[scale=0.7]{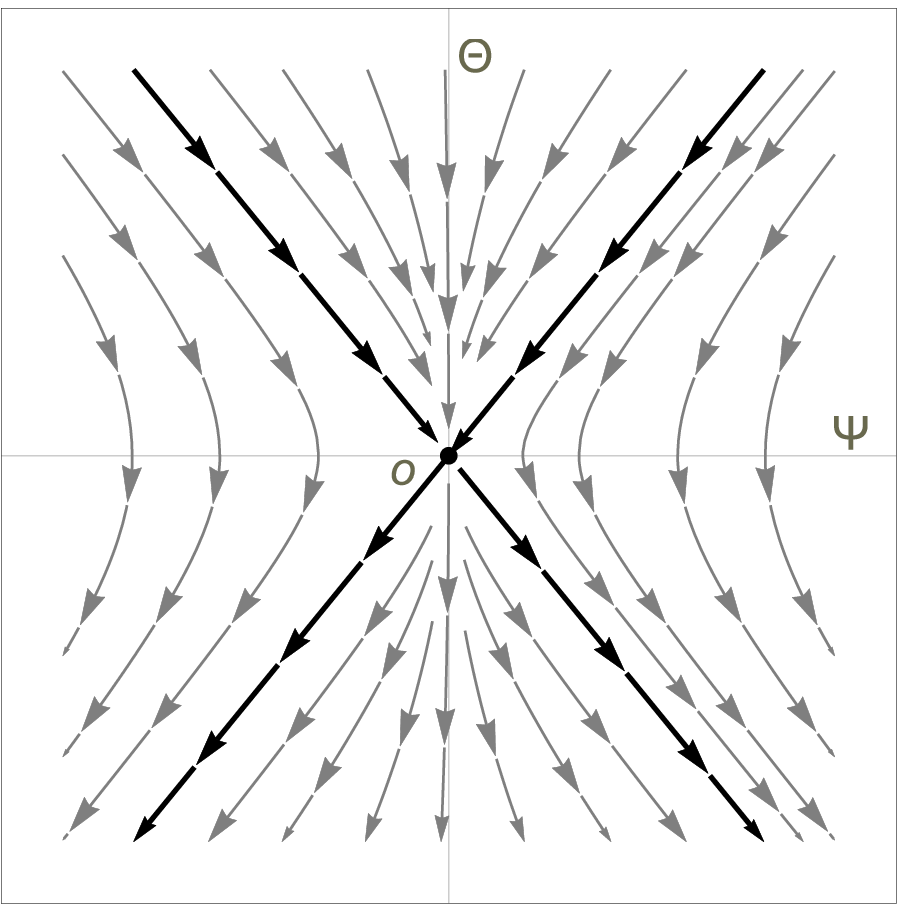}}\caption{Phase portrait for
cosmological constant potential}%
\end{figure}\FloatBarrier
In order to draw the phase portrait, let us first determine the critical (or
equilibrium) points of the system, that is, the points in the phase plane at
which the right side of (\ref{poster16}) and (\ref{poster17}) vanishes. There
are four equilibrium points, but only two of them satisfy the constraint
equation (\ref{poster15}), and these correspond to special solutions,
presenting the simplest kind of behaviour. It is not difficult to see that we
have equilibrium points only if $\Lambda>0$ \footnote{Clearly, the case when
$\omega$ and $\Lambda$ are both negative is not allowed because of
(\ref{poster15}).}. These are $A$ $(0,\sqrt{\frac{3\Lambda}{2}})$ and $B$
$(0,-\sqrt{\frac{3\Lambda}{2}})$, which lie on the axis $\psi=0,$ and
correspond, respectively, to the solutions (\ref{cosmological_const01}), i.e.
, the de Sitter and anti-de Sitter solutions found in the previous section.
\ The solutions for $\omega<0$ and $\Lambda>0$, obtained above, lie in the
ellipse (depicted in bold), which is the only curve in the phase plane
satisfying the constraint (\ref{poster15}) (see \textbf{Fig. 1a}). The
solutions given by the equations (\ref{cosmological_const02}) and
(\ref{cosmological_const03}) are represented in the diagram as the
trajectories starting at $B$ and ending at $A$, in the anticlockwise
$(BA^{+})$ and clockwise sense $(BA^{-})$, \ respectively. \ It is interesting
to note that the solution corresponding to the point $A$ is stable with
respect to small perturbations in the phase plane, while $B$ has a unstable
character. The existence of this kind of stability (instability) pattern
associated to the two critical points seems to lead to an interesting
behaviour as far as the space-time geometry of these models is concerned.
Consider, for instance, the solution represented by the equilibrium point $B$,
which, as we know, corresponds to a contracting de Sitter space-time, with a
constant scalar field. From the point of view of the Weyl frame $(M,g,\phi)$,
this means that the space-time geometry is Riemannian. Now, let us look at how
the space-time geometry evolves when this universe is slightly perturbed.
\ Clearly, the time evolution is either determined by $BA^{+}$ or $BA^{-}$ .
Initially, when $t\rightarrow-\infty$, these universes are in an purely
Riemannian regime as the scalar field is nearly constant. This situation will
change as soon as the scalar field ceases to be a constant and gradually
starts to depend on time. Then\ the universe enters a new regime, in which
space-time is characterized by a Weyl integrable geometry. One is tempted to
say that we have here a kind of geometrical phase transition, since we go from
a purely Riemannian geometry to a Weyl integrable geometry. Finally, as
$t\rightarrow\infty$ , since $BA^{+}$ and $BA^{-}$\ both approach the
equilibrium point $A$, the universe returns to its initial Riemannian regime,
undergoing another geometrical transition, this time from a Weyl geometry to a
purely Riemannian de Sitter space-time, represented in the phase portrait by
the equilibrium point $A$.

The phase diagram corresponding to the solutions for $\omega>0$ and
$\Lambda<0$, namely, (\ref{cosmological_const04}) and
(\ref{cosmological_const05}), are depicted as bold lines in \textbf{Fig. 1b}.
As the diagram clearly shows, we have two singular\ universes which start with
a big bang, undergo an era of expansion, and finally collapse to a big crunch.
Interpreting this picture in the Weyl frame, we see that because the time
derivative of the scalar field does not vanish no geometrical phase transition
takes place in this case.

Finally, when both $\omega$ and $\Lambda$ are positive, the phase portrait of
the solutions is shown in \textbf{Fig. 1c}. In this case, as regards the
solutions given by equations (\ref{cosmological_const06}) and
(\ref{cosmological_const07}), the critical point $A$ behaves as an attractor,
while $B$ acts as a repellor. In the first situation, we have singular
expanding Weyl space-times evolving towards a Riemannian de Sitter universe,
while in the second, small perturbations cause a Riemannian anti-de Sitter
universe start to collapsing into a big crunch.

To conclude this section, let us, just for the sake of completeness, examine
the degenerate case when $\Lambda=0$. In this case, the equations
(\ref{poster12}), (\ref{poster13}) and (\ref{poster14}) \ become
\begin{equation}
\frac{\theta^{2}}{3}=\frac{\omega}{2}\psi^{2} \label{nolambda1}%
\end{equation}%
\begin{equation}
\dot{\theta}=-\frac{1}{2}\theta^{2}-\frac{3\omega}{4}\psi^{2}
\label{nolambda2}%
\end{equation}%
\begin{equation}
\dot{\psi}=-\theta\psi. \label{nolambda3}%
\end{equation}

We first note that in order to obtain real solutions we \ must restrict the
parameter $\omega$ to be positive \footnote{The case $\omega=0$ is trivial,
corresponding to Minkowski space-time.}. We also note that the conics that
appeared in the former phase diagrams, representing the constraint of the
dynamical system, now simply degenerate into the pair of straight lines
$\theta=\pm\sqrt{\frac{3\omega}{2}}\psi$ (see \textbf{Fig. 1d}). \ The
equations above are easily integrated and give the following solutions:%

\begin{equation}
a(t)=a_{0}|t-t_{0}|^{\frac{1}{3}}, \label{a}%
\end{equation}%
\begin{equation}
\phi(t)=\phi_{0}\pm\sqrt{\frac{2}{3\omega}}\ln\left\vert t-t_{0}\right\vert .
\label{fi}%
\end{equation}
We now have only one critical point, which lies at the origin $\mathcal{O}$,
and this clearly corresponds to Minkowski space-time. For $t>t_{0}$ the two
solutions given by (\ref{a}) and (\ref{fi}) start with a big bang and approach
$\mathcal{O}$ as $t\rightarrow\infty$ . On the other hand, for $t<t_{0}$ we
have two other solutions which comes from $\mathcal{O}$ when $t\rightarrow
-\infty$, and then collapses to a singular space-time as $t\rightarrow t_{0}$.

Finally, it should be remarked here that when $\omega<0$, the action
(\ref{action}) includes a phantom field. It is known that in this case we have
violation of the null energy condition \cite{Rubakov1}. (It is important to
note that the case of constant potential and $\omega<0$, has been investigated
in a different context (mainly inspired by string field theories) in
\cite{Arefeva1}.

Let us now briefly consider other types of scalar potentials.

\subsection{Potential of a massive scalar field}

In this section, we shall briefly consider a potential of the type
$V(\phi)=e^{-2\phi}(m^{2}\phi^{2}+\Lambda).$ In the Riemann frame, this type
of potential, which corresponds to the case of a massive scalar field plus a
cosmological constant is easily found in the literature of inflationary
models, and leads to several different cosmological regimes \cite{Felder}.

By applying a known simple mathematical procedure (first-order formalism) to
the field equations (\ref{poster12})-(\ref{poster14}) we obtain the following
solution \cite{Bazeia}%

\begin{equation}
\phi(t)=\phi_{0}+3\frac{\alpha\Lambda}{m^{2}}(t-t_{0}), \label{poster27}%
\end{equation}%
\begin{equation}
a(t)=a_{0}\exp\left(  \alpha\phi_{0} (t-t_{0})+\frac{\Lambda}{4}(t-t_{0}%
)^{2}\right)  , \label{poster26}%
\end{equation}
where
\begin{equation}
\alpha^{2}=\frac{m^{2}}{6},\quad\Lambda=-\frac{2m^{2}}{3\omega},
\label{poster28}%
\end{equation}
and $a_{0}$, $t_{0}$ and $\phi_{0}$ are constants of integration.

The field equations for this potential may be put in the form
\begin{equation}
\frac{\theta^{2}}{3}=\frac{\omega}{2}\psi^{2}+\frac{1}{2}(m^{2}\phi
^{2}+\Lambda), \label{mp2}%
\end{equation}%
\begin{equation}
\dot{\theta}=-\frac{\theta^{2}}{2}-\frac{3\omega}{4}\psi^{2}+\frac{3}{4}%
(m^{2}\phi^{2}+\Lambda), \label{mp3}%
\end{equation}%
\begin{equation}
\dot{\psi}=-\theta\psi-\frac{m^{2}}{\omega}\phi. \label{mp4}%
\end{equation}
In the above equations, we can use (\ref{mp2}) to eliminate $\phi$ by writing
$\phi=\pm\frac{1}{m}\sqrt{\frac{2}{3}\theta^{2}-\omega\psi^{2}-\Lambda}$ , and
thus arrive at the following dynamical system, defined only in terms of the
variables $\theta$ and $\psi$:%

\begin{equation}
\label{poster29}\dot{\theta}= -\frac{3}{2} \omega\psi^{2},
\end{equation}

\begin{equation}
\label{poster30}\dot{\psi}=-\theta\psi\mp\frac{m^{2}}{\omega} \sqrt{\frac
{1}{m^{2}}\left(  \frac{2}{3}\theta^{2}-\omega\psi^{2}-\Lambda\right)  }.
\end{equation}
Actually, we have two dynamical systems according to whether we take $+$ or
$-$ in (\ref{poster30}). The phase portrait of the solutions is displayed
below. \FloatBarrier
\begin{figure}[tbh]
\centering
\subfloat[ $\alpha (t-t_0)> \frac{\omega \phi_0}{2}$]{
\includegraphics[scale=0.7]{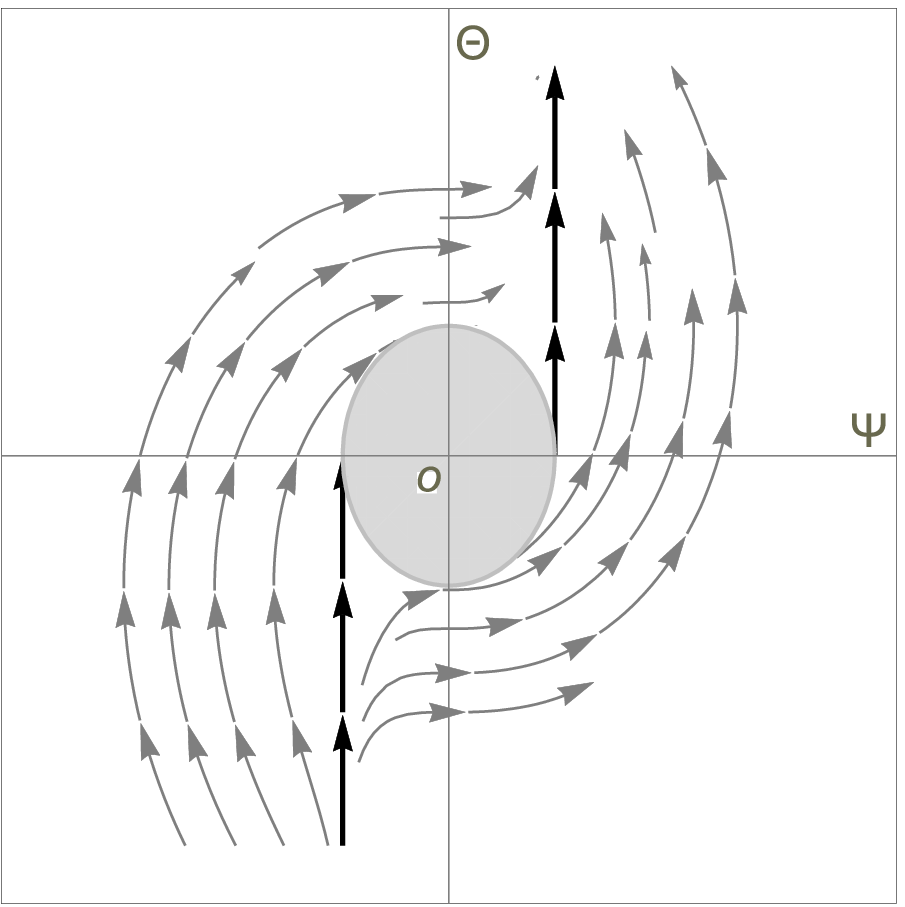}
} \quad\subfloat[ $\alpha (t-t_0)<\frac{\omega \phi_0}{2}$]{
\includegraphics[scale=0.7]{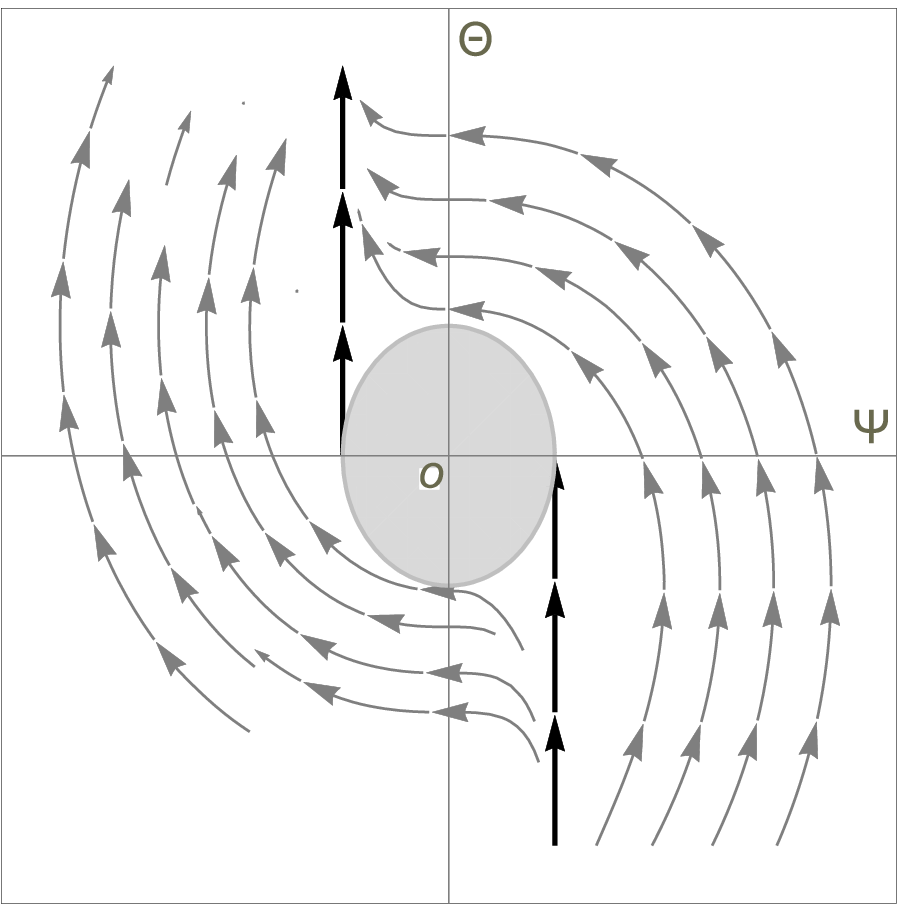}
}\caption{Phase portraits for massive scalar field potential}%
\end{figure}\FloatBarrier
Let us now make some comments on the behaviour of the solutions. The critical
points are given by $\psi=0$ and $\theta=\pm\sqrt{\frac{3}{2}\Lambda}$ , and,
although they are solutions of the dynamical system defined by (\ref{poster29}%
) and (\ref{poster30}), they do not represent a solution of the complete set
of field equations for $\Lambda\neq0$. The physical solutions (\ref{poster27})
and (\ref{poster26}) correspond to the isoclines $\psi=\pm\sqrt{\frac
{2m}{3|\omega|}}$, and are represented by the two bold straight lines in the
diagrams (see \textbf{Fig. 2}, above). It is to be noted that these solutions
are continuous with respect to the time parameter $t$, and that, as time goes
by, they pass from one diagram to the other diagram continuously. Clearly, the
two solutions describe non-singular universes that undergo a contraction era
followed by a expanding period, depending on the sign we ascribe to the
constant $\alpha$. It also should be noted that, as can be seen from
(\ref{poster30}), in the (shaded) elliptic region bounded by the curve
$\frac{2}{3}\theta^{2}-\omega\psi^{2}=\Lambda$ the dynamical system is not
defined. \textbf{Fig. 2} shows the phase portrait for $\omega<0$ and
$\Lambda>0$, satisfying the condition (\ref{poster28}). Let us mention here
that models with quadratic potential, appearing in a different context, have
been previously considered in which the phase portraits corresponding to the
field equations are also discussed \cite{Arefeva2}.

\subsection{Exponential scalar potential}

Some well-known inflationary models assume that the evolution of the universe
during inflation is driven by a scalar field generated by an exponential
potential of the form
\begin{equation}
V(\phi)=V_{0}e^{-(\lambda+2)\phi}, \label{exp_pot01}%
\end{equation}
with $V_{0}$ and $\lambda$ $(\lambda>0)$ being constants \cite{Lucchin}. It is
not difficult to verify that, in this case, the field equations
(\ref{poster12}), (\ref{poster13}) and (\ref{poster14}) have the following
solutions:
\begin{equation}
a(t)=a_{0}\left(  \pm\frac{\lambda^{2}}{2}\sqrt{\frac{V_{0}}{\omega
(6\omega-\lambda^{2})}}\;e^{-\frac{\lambda}{2}\phi_{0}}(t-t_{0})+1\right)
^{2\omega/\lambda^{2}}, \label{exp_pot02}%
\end{equation}%
\begin{equation}
\phi(t)=\frac{2}{\lambda}\ln\bigg|\pm\frac{\lambda^{2}}{2}\sqrt{\frac{V_{0}%
}{\omega(6\omega-\lambda^{2})}}\;(t-t_{0})+e^{\frac{\lambda}{2}\phi_{0}%
}\bigg|, \label{exp_pot03}%
\end{equation}
Let us remark that these solutions are in agreement with the already known
result that exponential potentials generate power-law inflation. (Note that
the possible values of the free parameter $\omega$ are restricted to the
intervals $\omega>\frac{\lambda^{2}}{6}$ and $\omega<0$.) On the other hand,
the expansion parameter of the model is given by
\begin{equation}
\theta(t)=\theta_{0}e^{-\frac{\lambda}{2}\phi_{0}}\left(  \pm\frac{\lambda
^{2}}{2}\sqrt{\frac{V_{0}}{\omega(6\omega-\lambda^{2})}}\;e^{-\frac{\lambda
}{2}\phi_{0}}(t-t_{0})+1\right)  ^{-1}.
\end{equation}
To get a clearer picture of the behaviour of the solutions let us examine the
dynamical system obtained from the field equations. It is not difficult to
verify that from the set of equations (\ref{eq01}), (\ref{poster31}),
(\ref{eq3}) reduces to
\begin{equation}
\frac{\theta^{2}}{3}=\frac{\omega}{2}\psi^{2}+\frac{V_{0}}{2}e^{-\lambda\phi}
\label{exp1}%
\end{equation}%
\begin{equation}
\dot{\theta}=\frac{\theta^{2}}{2}-\frac{3}{4}\omega\psi^{2}+\frac{3}{4}%
V_{0}e^{-\lambda\phi}, \label{exp2}%
\end{equation}%
\begin{equation}
\dot{\psi}=-\theta\psi+\frac{\lambda}{2\omega}V_{0}e^{-\lambda\phi}.
\label{exp3}%
\end{equation}
With the help of (\ref{exp1}) the equations (\ref{exp2}) and (\ref{exp3}) may
be written as
\begin{equation}
\dot{\theta}=-\frac{3}{2}\omega\psi^{2}, \label{exp4}%
\end{equation}%
\begin{equation}
\dot{\psi}=-\theta\psi+\frac{\lambda}{3\omega}\theta^{2}-\frac{\lambda}{2}%
\psi^{2}. \label{exp5}%
\end{equation}
\FloatBarrier
\begin{figure}[tbh]
\centering
\subfloat[$\omega> \frac{\lambda^2}{6}$]{
\includegraphics[scale=0.7]{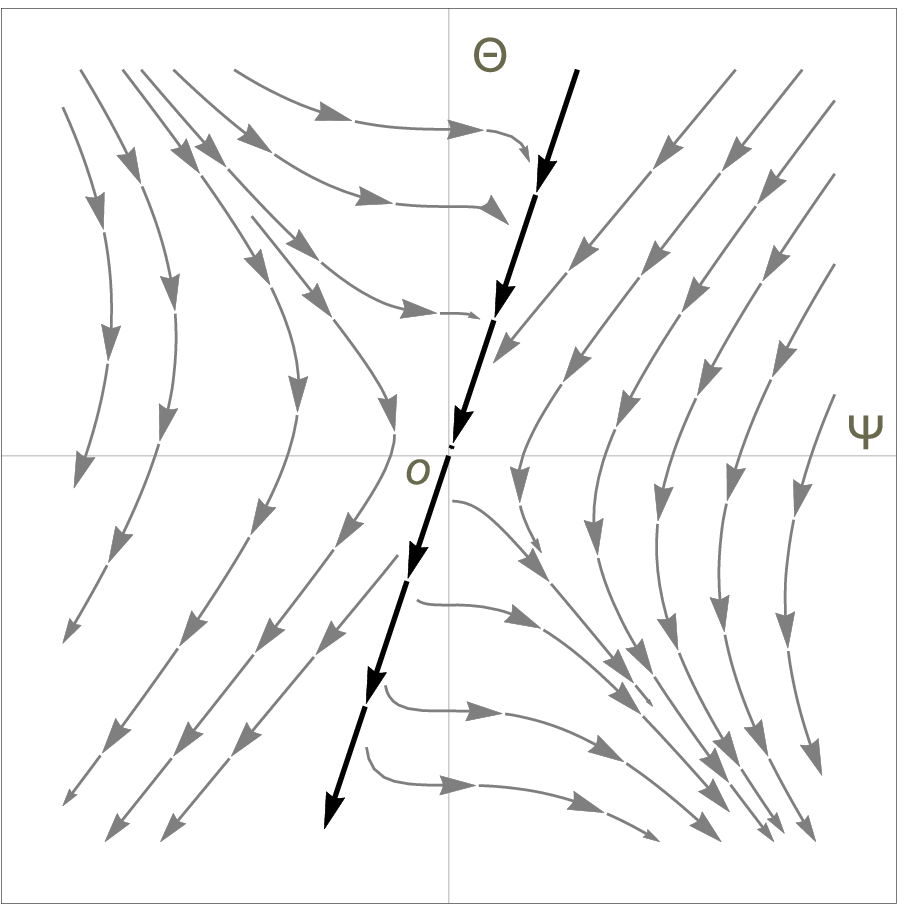}
} \quad
\subfloat[$\omega<0$]{
\includegraphics[scale=0.7]{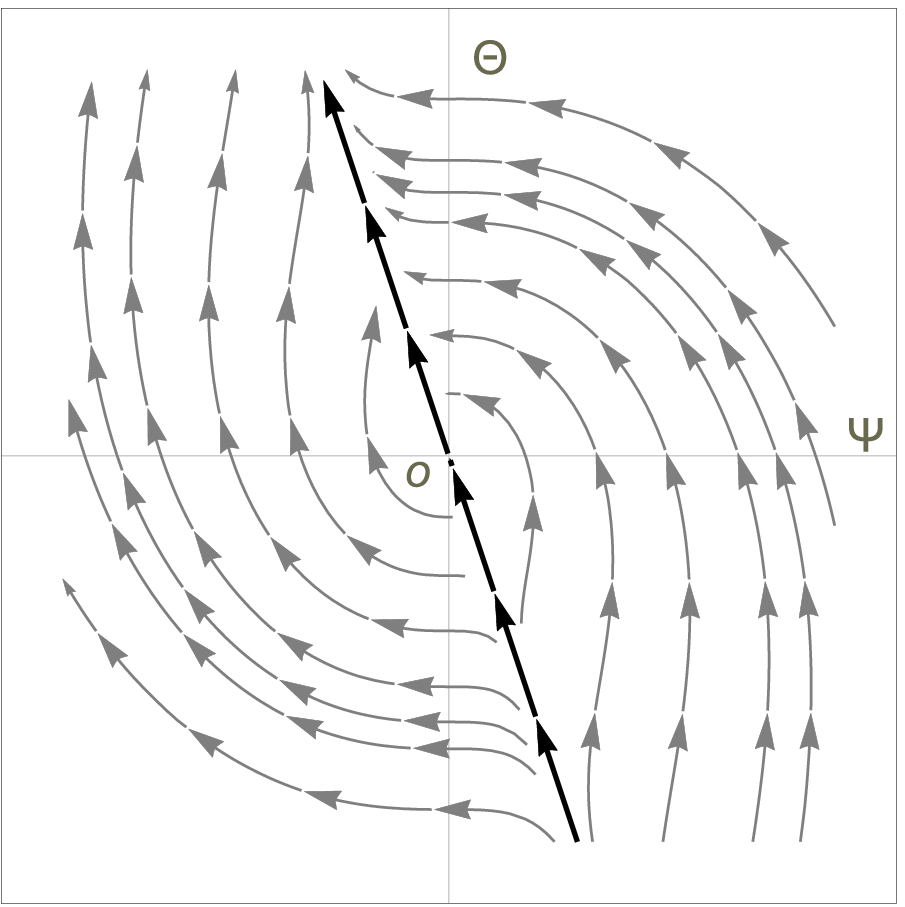}
}\caption{Phase portrait for exponential scalar potential}%
\end{figure}\FloatBarrier
The phase portrait of this dynamical system is displayed in \textbf{Fig. 3},
where the solutions given by (\ref{exp_pot02}) and (\ref{exp_pot03}), lie on
the straight line $\theta=\frac{3\omega}{\lambda}\psi$, passing through the
origin $\mathcal{O}$, which is the only equilibrium point of the system, and
corresponds to Minkowski space-time. In \textbf{Fig. 3a}, we have depicted the
solutions for $\omega>\frac{\lambda^{2}}{6}$, which represent an expanding
model singular, approaching Minkowski space-time as $t\rightarrow\infty$, and
a collapsing model starting from Minkowski space-time at $t\rightarrow-\infty$
evolving towards a singularity.

\subsection*{Quartic potential}

We next consider the quadratic potential
\begin{equation}
V(\phi)=2\lambda(\phi^{2}-\beta)^{2}e^{-2\phi}, \label{scalar-tensor34}%
\end{equation}
where $\lambda$ and $\beta$ are positive constants. This particular kind of
\textit{effective} quartic potential has been considered in inflationary
scenarios mainly inspired in the idea that it is the Higgs boson that plays
the role of the inflaton field \cite{Bezrukov}.

We can easily show that the cosmological equations (\ref{poster12}),
(\ref{poster13}) and (\ref{poster14}) admit the following solution:
\begin{equation}
\label{scalar-tensor29}\phi(t)=\phi_{0}\exp\left(  -\frac{4A}{\omega}%
(t-t_{0})\right)  ,
\end{equation}%
\begin{equation}
\label{scalar-tensor30}a(t)=a_{0}\exp\left\{  -\frac{\omega{\phi_{0}}^{2}}%
{8}\left[  \exp\left(  -\frac{8A}{\omega}(t-t_{0})\right)  -1\right]
+B(t-t_{0})\right\}  ,
\end{equation}
with $\phi_{0}$, $a_{0}$ being constants of integration, $A^{2}=\frac{\lambda
}{3},$ $B^{2}=\beta^{2}A^{2}$, and the condition $\beta=\frac{2}{3\omega}$
must be satisfied. The expansion factor gives
\begin{equation}
\theta(t)=3B+3A{\phi_{0}}^{2}\exp\left(  -\frac{8A}{\omega}(t-t_{0})\right)  .
\end{equation}
Clearly, these correspond to non-singular universes undergoing expansion or
contraction, depending on the value assumed by the constants.

If we wish to treat the field equations for this potential as a dynamical
system, we write them in the form
\begin{equation}
\frac{\theta^{2}}{3}=\frac{\omega}{2}\psi^{2}+\lambda(\phi^{2}-\beta)^{2},
\label{constraint quart pot}%
\end{equation}%
\begin{equation}
\dot{\theta}=-\frac{\theta^{2}}{2}-\frac{3\omega}{4}\psi^{2}+\frac{3\lambda
}{2}(\phi^{2}-\beta)^{2}, \label{eq motion quart pot}%
\end{equation}%
\begin{equation}
\dot{\psi}=-\theta\psi-\frac{4\lambda}{\omega}\phi(\phi^{2}-\beta).
\end{equation}
As in the case of the massive scalar field, the constraint equation
(\ref{constraint quart pot}) can be used to eliminate the variable $\phi$ from
the dynamical equations. This procedure will lead us to four distinct
dynamical systems. These are given by the following equations:
\begin{equation}
\dot{\theta}=-\frac{3}{2}\omega\psi^{2}, \label{sist din quartic pot theta}%
\end{equation}%
\begin{equation}
\dot{\psi}=-\theta\psi-\left[  \pm\frac{4\lambda}{\omega}\sqrt{\pm\sqrt
{\frac{1}{\lambda}\left(  \frac{\theta^{2}}{3}-\frac{\omega}{2}\psi
^{2}\right)  }+\beta}\right]  \left[  \pm\sqrt{\frac{1}{\lambda}\left(
\frac{\theta^{2}}{3}-\frac{\omega}{2}\psi^{2}\right)  }\right]  .
\label{sist din quartic pot psi}%
\end{equation}
The first pair of signs $\pm$ in the right-hand side of
(\ref{sist din quartic pot psi}) defines two dynamical systems, corresponding
to the two possibles signs of $\phi$. The other two dynamical systems arise
when the second and third pairs of plus or minus signs are fixed simultaneously,
according to whether $\phi^{2}-\beta>0$ or $\phi^{2}-\beta<0$. The phase
portraits of these four dynamical systems (\ref{sist din quartic pot theta}%
)-(\ref{sist din quartic pot psi}) are depicted in Figure 4, below.

\FloatBarrier\begin{figure}[tbh]
\centering
\subfloat[$\phi>0$ and $\phi^2-\beta>0$]{\includegraphics[scale=0.7]{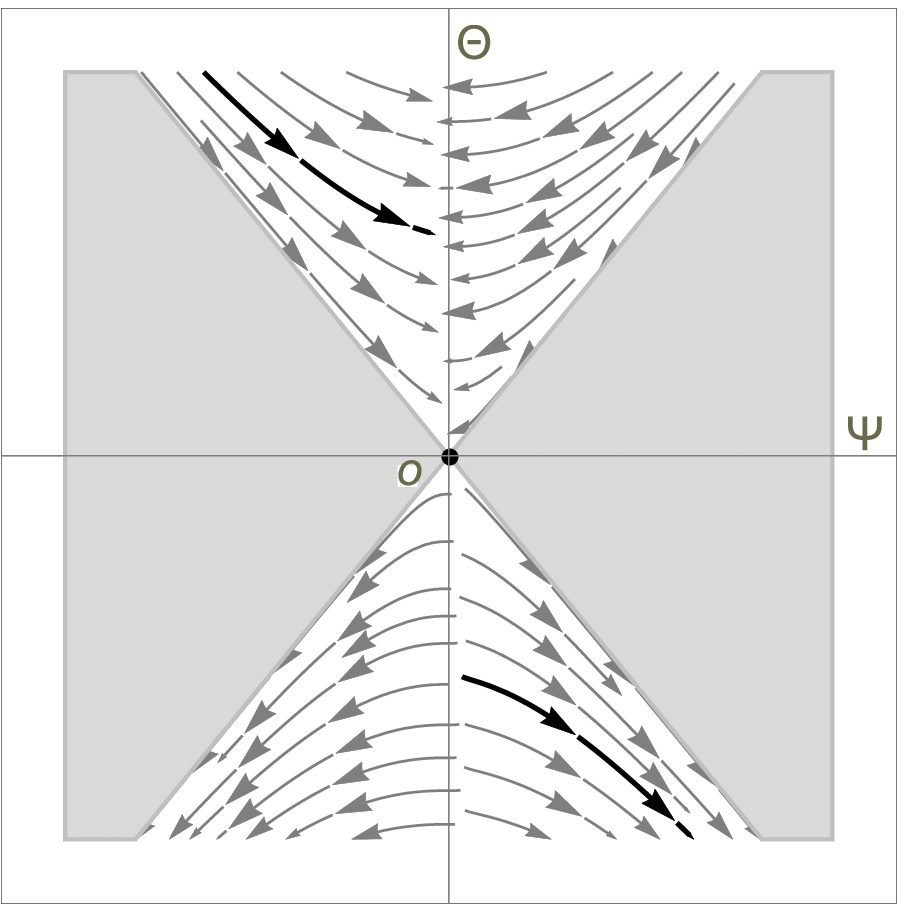}}
\quad
\subfloat[$\phi>0$ and $\phi^2-\beta<0$]{\includegraphics[scale=0.7]{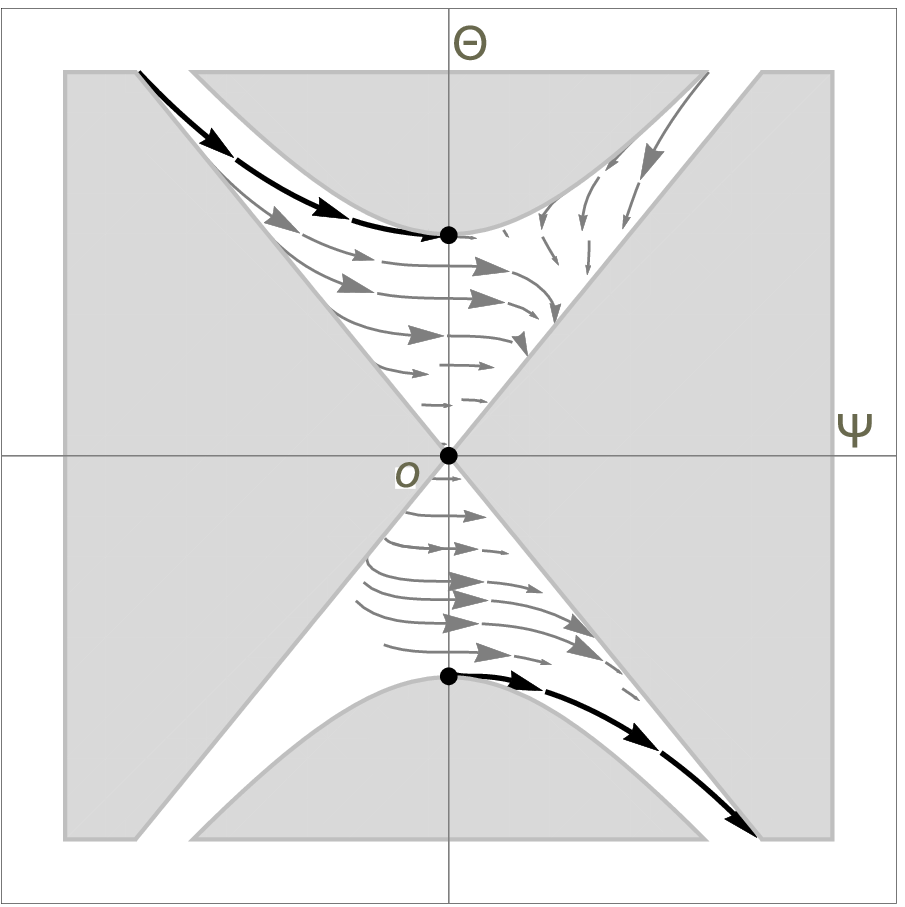}}
\quad
\subfloat[$\phi<0$ and $\phi^2-\beta>0$]{\includegraphics[scale=0.7]{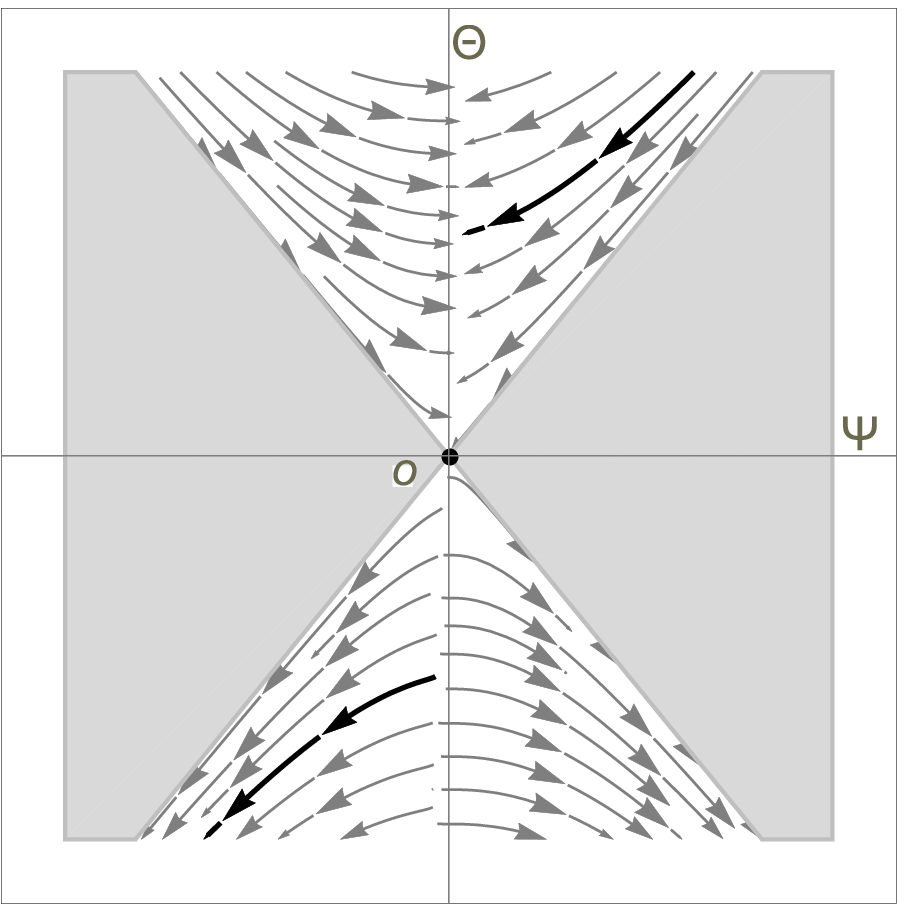}}
\quad
\subfloat[$\phi<0$ and $\phi^2-\beta<0$]{\includegraphics[scale=0.7]{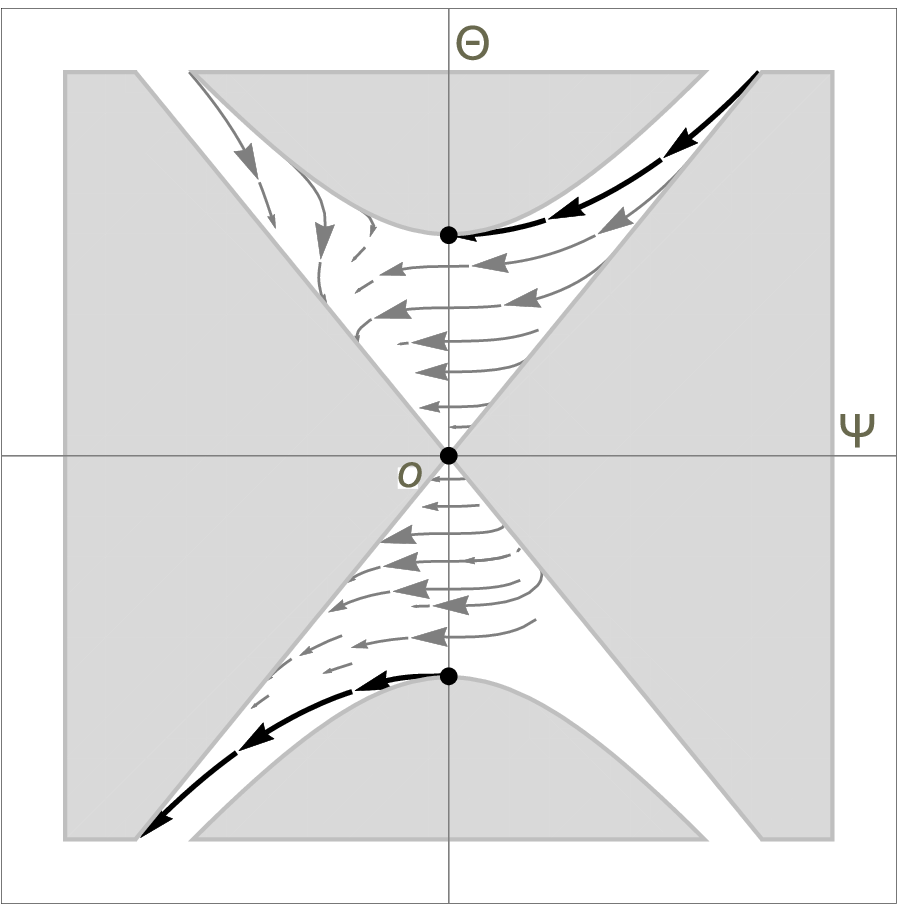}}\caption{Phase
portrait for quartic potential}%
\end{figure}\FloatBarrier

As to the critical points, it is not difficult to verify that the origin is an
equilibrium point of the four dynamical systems
(\ref{sist din quartic pot theta})-(\ref{sist din quartic pot psi}), which
corresponds to the trivial case of Minkowski space-time with a constant scalar
field $\phi=\pm\sqrt{\beta}$ (see Eq. (\ref{constraint quart pot})). On the
other hand, the two dynamical systems for which $\phi^{2}-\beta<0$ have two
additional critical points: $(0,\pm\beta\sqrt{3\lambda})$. These represent
solutions of the field equations, corresponding to an expanding or a
contracting de Sitter universe with a null scalar field.

A comment on the domain of the phase plane where the dynamical systems are
defined is now in order. Clearly, the square roots in
(\ref{sist din quartic pot theta})-(\ref{sist din quartic pot psi}) restrict
the possible values of the dynamical variables. In all cases, the inequality
$\frac{\theta^{2}}{3}\geq\frac{\omega\psi^{2}}{2}$ must be satisfied, while in
two of them (\textbf{Fig. 4b} and \textbf{Fig. 4d}) we have an additional restriction imposed by
$\sqrt{\frac{1}{\lambda}\left(  \frac{\theta^{2}}{3}-\frac{\omega}{2}\psi
^{2}\right)  }\leq\beta$.

It is not difficult to verify that the analytic solutions
(\ref{scalar-tensor29})-(\ref{scalar-tensor30}) lie on the parabola
$\theta=3B+\frac{3\omega^{2}}{16A}\psi^{2}$, where we are considering the case
in which the constants $A$ and $B$ have the same sign. As an example, let us
take $A>0$ and $\phi_{0}>0$, which in turns implies that $\psi<0$. In this
case, the analytic solution will be represented by the curve (in bold) shown
in the second quadrant $(\theta>0,$ $\psi<0)$ of \textbf{Fig. 4a} and \textbf{Fig. 4b}.
Moreover, it is the sign of $\phi^{2}-\beta$ that determines the time interval
for which this curve is a solution of the field equations. For instance, if
$\phi^{2}-\beta>0$, then $t-t_{0}<-\frac{\omega}{8A}\ln|\beta/\phi_{0}^{2}|$ ,
and, thus, the bold line in \textbf{Fig. 4a} represents the analytic solution in that
interval. For $t-t_{0}>-\frac{\omega}{8A}\ln|\beta/\phi_{0}^{2}|$ , we must
look at the diagram of \textbf{Fig. 4b}, where $\phi^{2}-\beta<0$. Here, the solution
is represented by the curve (in bold) approaching the critical point
$(0,\beta\sqrt{3\lambda})$ as $t\rightarrow\infty$. Similarly, the analytic
solutions (\ref{scalar-tensor29})-(\ref{scalar-tensor30}) for negative $\phi$
appear in \textbf{Fig. 4c} and \textbf{Fig. 4d}, where again the time interval of validity is
determined by the sign of $\phi^{2}-\beta$. Finally, let us mention that the
curves lying in the region $\theta<0$ of all the diagrams correspond to the
choice $A<0$ and $B<0$, with a continuous dependence on time going from one
diagram to another and approaching the critical point $(0,-\beta\sqrt
{3\lambda}).$ To conclude, let us note that the critical point $(0,\beta
\sqrt{3\lambda})$ is a stable solution for this model, whereas the critical
point $(0,-\beta\sqrt{3\lambda})$ is unstable.

\section{A cosmological toy model with non-singular behaviour and geometric
phase transition}

As we have previoulsly mentioned, in the last two or three decades there has
been a great deal of work on the inflationary program, as well as in dark
energy models, in which the scalar field plays a vital role \cite{Arefeva3}.
However, the fact that the nature of the scalar field which is supposed to
drive the inflationary process or accelerate the universe is not yet known may
lead us to conjecture whether one could attribute a pure geometric character
to this field. We shall not attempt here to examine this question, which we
leave for future research. Instead, in this section, we shall briefly sketch a
very simple model, say, a \textquotedblleft toy model\textquotedblright, that
seems to exhibit in a rough qualitative way some interesting\ features of a
pure geometric scalar-tensor model. In particular, we have found a
cosmological scenario which might be viewed as qualitatively describing a kind
of geometric phase transition of the universe. It is to be noted,
incidentally, that for $\omega<0$ we have a phantom scalar field. Models of
this kind are already known and have been recently investigated by some
authors\ to describe dark energy using string motivated models \cite{Arefeva3}.

We shall start with the following power-law potential:
\begin{equation}
V(\phi)=6e^{-2\phi}\left[  \alpha-\frac{\omega\beta}{6}\left(  3 \sigma
\phi-\frac{\phi^{3}}{\sigma}\right)  \right]  ^{2}-\omega\sigma^{2}\beta
^{2}e^{-2\phi}\left(  1-\frac{\phi^{2}}{\sigma^{2}}\right)  ^{2}, \label{eq04}%
\end{equation}
where $\alpha$ is a positive constant, $\beta$ and $\sigma$ are arbitrary
constants. The field equations (\ref{eq01}), (\ref{poster31}) and (\ref{eq3})
take the form
\begin{equation}
\frac{\theta^{2}}{3}=\frac{\omega}{2}\psi^{2}+3\left[  \alpha-\frac
{\omega\beta}{6}\left(  3 \sigma\phi-\frac{\phi^{3}}{\sigma}\right)  \right]
^{2}-\frac{1}{2}\omega\sigma^{2}\beta^{2}\left(  1-\frac{\phi^{2}}{\sigma^{2}%
}\right)  ^{2}, \label{eq05}%
\end{equation}%
\begin{equation}
\dot{\theta}=-\frac{\theta^{2}}{2}-\frac{3\omega}{4}\psi^{2}+\frac{9}%
{2}\left[  \alpha-\frac{\omega\beta}{6}\left(  3 \sigma\phi-\frac{\phi^{3}%
}{\sigma}\right)  \right]  ^{2}-\frac{3}{4}\omega\sigma^{2}\beta^{2}e^{-2\phi
}\left(  1-\frac{\phi^{2}}{\sigma^{2}}\right)  ^{2}, \label{eq06}%
\end{equation}%
\begin{equation}
\dot{\psi}=-\theta\psi+3\beta\sigma\left(  1-\frac{\phi^{2}}{\sigma^{2}%
}\right)  \left[  \alpha-\frac{\omega\beta}{6}\left(  3\sigma\phi-\frac
{\phi^{3}}{\sigma^{3}}\right)  \right]  -2\beta^{2}\phi\left(  1-\frac
{\phi^{2}}{\sigma^{2}}\right)  . \label{eq07}%
\end{equation}

Now, from Eqs. (\ref{eq05}) and (\ref{eq06}) it follows that
\begin{equation}
\theta(t)=3\alpha+\frac{3}{2}\sigma^{2}\beta\omega\left\{  \frac{1}{3}%
\tanh^{3}\left[  \beta(t-t_{0})\right]  -\tanh\left[  \beta(t-t_{0})\right]
\right\}  , \label{eq.08}%
\end{equation}
while the scalar factor and the scalar field are given by
\begin{equation}
a(t)=a_{0}\left\{  \tanh^{2}\left[  \beta(t-t_{0})\right]  -1\right\}
^{\frac{\sigma^{2}\omega}{6}}\exp\left\{  \beta(t-t_{0})-\frac{1}{12}%
\sigma^{2}\omega\tanh^{2}\left[  \beta(t-t_{0})\right]  \right\}  ,
\end{equation}%
\begin{equation}
\phi(t)=\sigma\tanh\left[  \beta(t-t_{0})\right]  . \label{fitan}%
\end{equation}
For specific choices of the values of the constants $\alpha$, $\beta$ and
$\sigma$, we can analyze the behaviour of the potential $V(\phi)$ and the
expansion parameter $\theta(t)$. A particularly interesting case,
corresponding to the choice $\alpha=-\frac{1}{3}\sigma^{2}\beta\omega$, with
$\omega<0$, is shown in \textbf{Fig. 5}, below. \FloatBarrier
\begin{figure}[th]
\subfloat[]{ \includegraphics[scale=0.7]{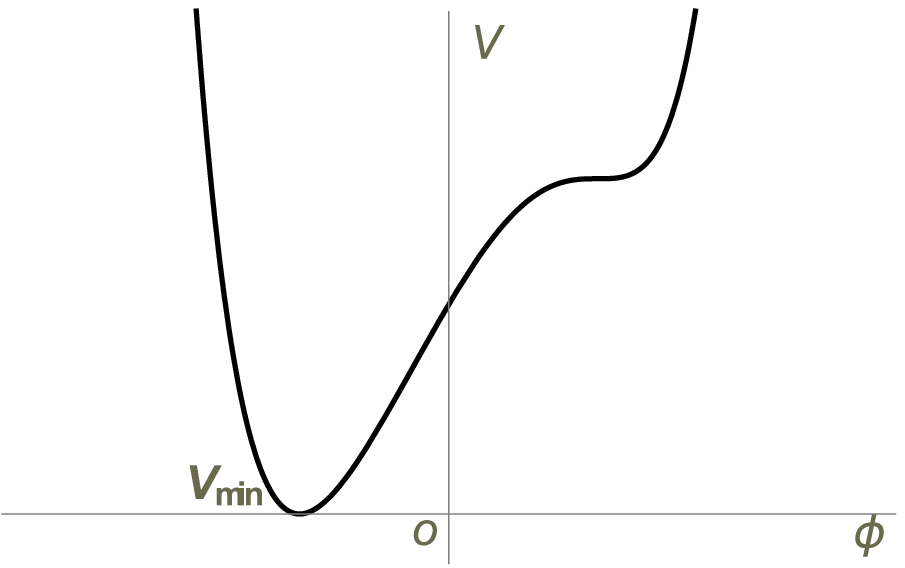}} \qquad
\subfloat[]{ \includegraphics[scale=0.7]{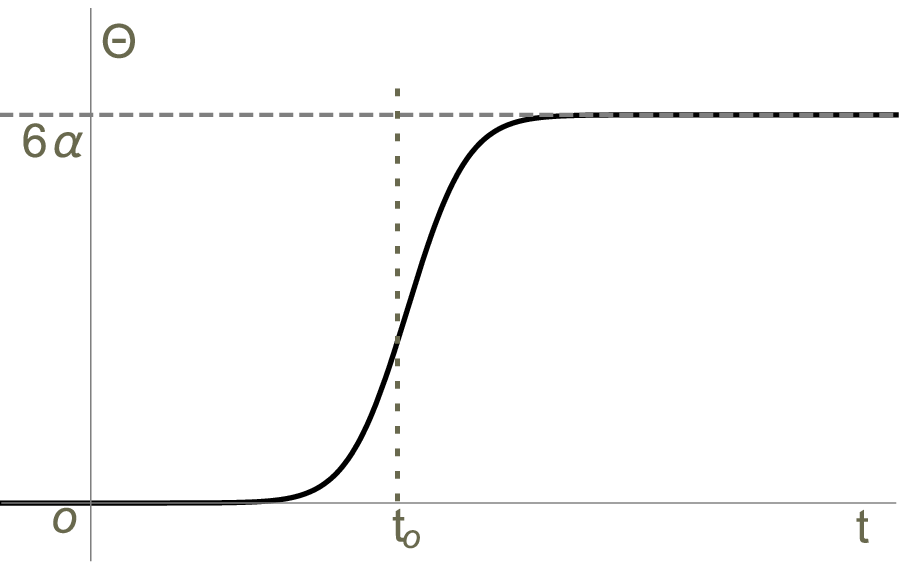}} \qquad\newline%
\subfloat[]{ \includegraphics[scale=0.7]{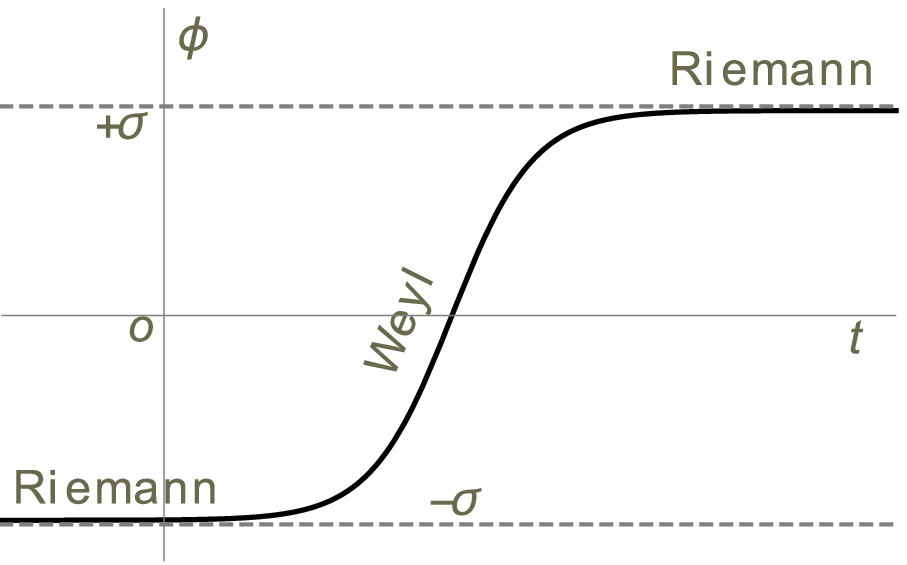}}
\qquad
\subfloat[]{ \includegraphics[scale=0.7]{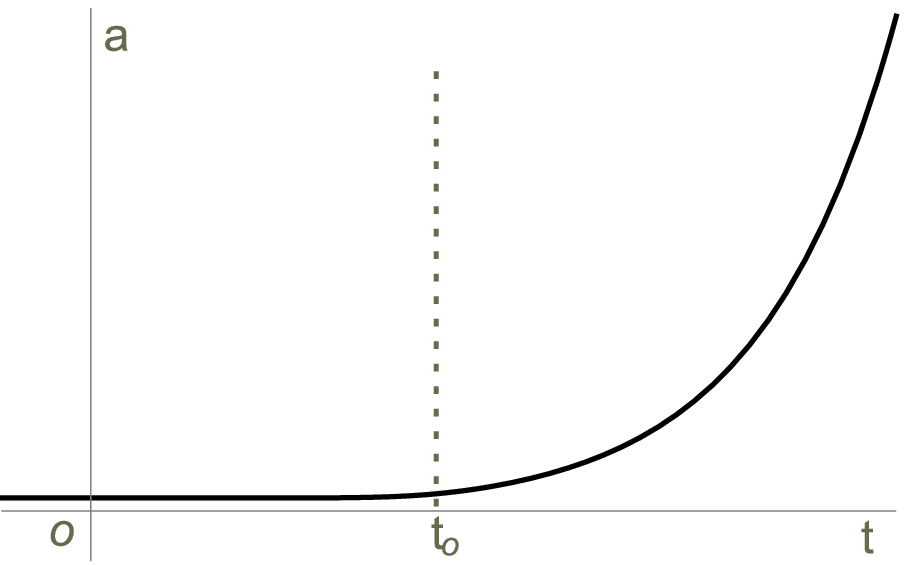}}\caption{Scalar
potential, expansion parameter, scalar field and scale factor for negative
$\omega$ and $\alpha\neq0$}%
\end{figure}\FloatBarrier
Let us briefly make some comments on this solution. From (\ref{fitan}) we see
that when $t\rightarrow\pm\infty$, $\phi$ tends asymptotically to $\pm\sigma$.
If we recall the geometrical meaning of the scalar field, we may interpret the
behaviour of the (kinklike) solution $\phi(t)$ as clearly indicating the
presence of two geometric phase transitions. Indeed, as we clearly see from
\textbf{Fig. 5c}, the universe comes asymptotically from a Riemannian regime
as $t\rightarrow-\infty$, undergoes a sudden expansion, and then goes back
smoothly (when $t\rightarrow+\infty$) to a Riemannian space-time. Although
these transitions are essentially continuous, we see that there is brief
period of time when the change in the space-time geometry is more drastic
\footnote{We would like to mention that by the time we were finishing the
present article we found that what we had called "geometrical phase
transition" was, in fact, already known, although in a slightly different
context. In the literature, the same phenomenon is referred to as a
"structural phase transition of the universe" \cite{Novello2}.}. This
coincides with the period when the expansion rate of the universe starts to
grow in a really significant way, taking much larger values than in the past
until it approaches a stage of exponential expansion. Clearly, the whole
expansion process is driven by the geometric scalar field $\phi$. In other
words, in this picture it is the dynamics of the scalar field \ that links the
quasi-static regime $(\theta\rightarrow0)$ to an expanding universe
assymptotically approaching a de Sitter regime ($\theta\rightarrow$ constant).
On the other hand, since in this model the de Sitter-like expansion phase of
the universe lasts forever.

Of course the above discussion is merely qualitative. Our aim in this work is
just to call the attention of cosmologists to new theoretical possibilities,
in which we can view the scalar field as possessing a pure geometric
character, being, in fact, part of the fundamental space-time structure.
Finally, according to this toy model, the universe is eternally existing, and
thus does not require a beginning or an ultimate end, and that means we have
here a simple example of an interesting dynamical cosmological scenario with
no singularity \cite{Ellis}.

To conclude, it is interesting to note that if we drop the above condition
$\alpha=-\frac{1}{3}\sigma^{2}\beta\omega$, we can, by appropriately choosing
the constants $\alpha$, $\beta$ and $\sigma$, obtain a class of the so-called
bounce models. As is well known, the general idea underlying the bounce
cosmology is that the hot big bang scenario, as it is understood today, simply
describes a period of expansion of the universe that followed a previous
contraction. In fact, a great deal of research has recently gone into the
study of these models \cite{Novello2, Biswas}. We therefore thought it would
be interesting to briefly mention that a bouncing universe scenario may also
emerge from a simple geometric model as the one presented in this section.
Indeed, it is not difficult to verify that for $\alpha=0$ the very same scalar
potential $V(\phi)$, given by (\ref{eq04}), leading to the equation
(\ref{eq.08}) allows for a non-singular bouncing universe whose behaviour is
displayed in the figures below. \FloatBarrier\begin{figure}[th]
\subfloat[]{ \includegraphics[scale=0.7]{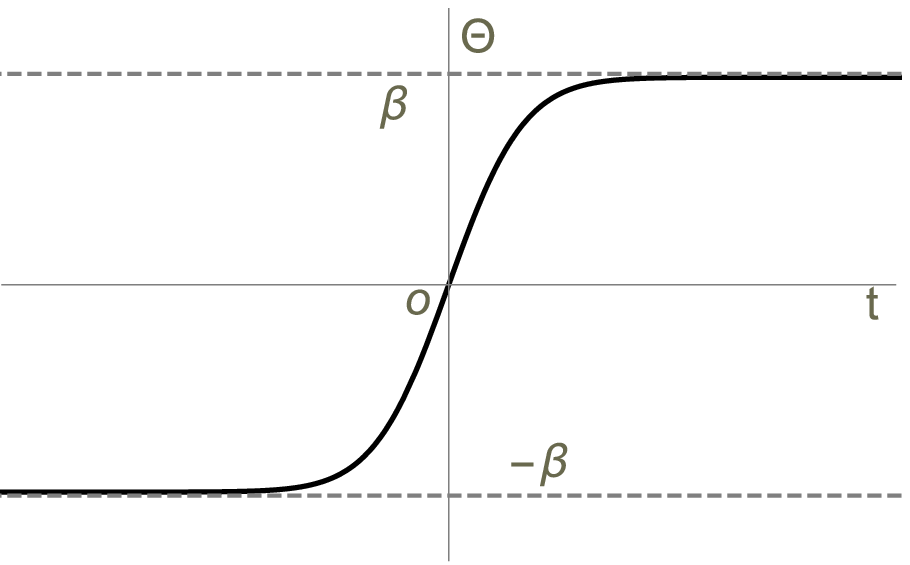}} \qquad
\subfloat[]{ \includegraphics[scale=0.7]{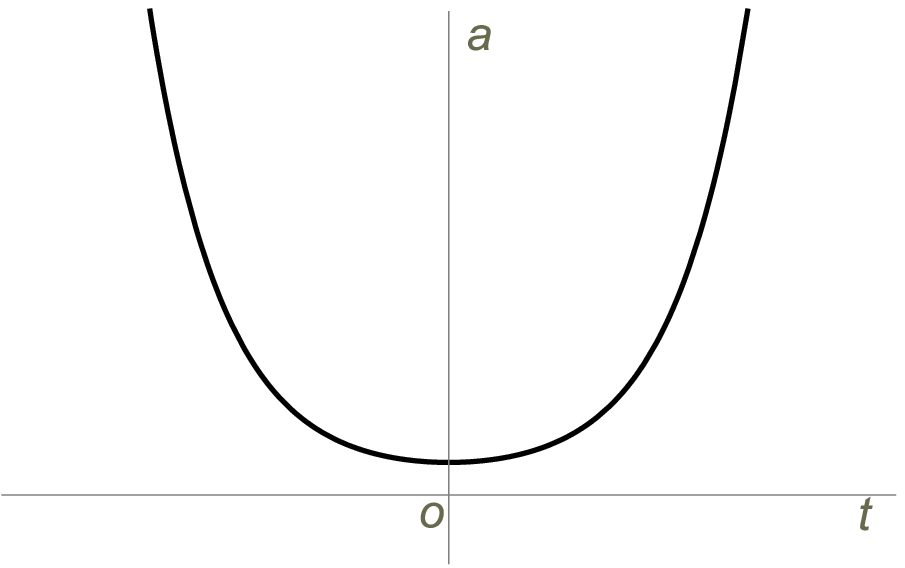}}
\qquad
\subfloat[]{ \includegraphics[scale=0.7]{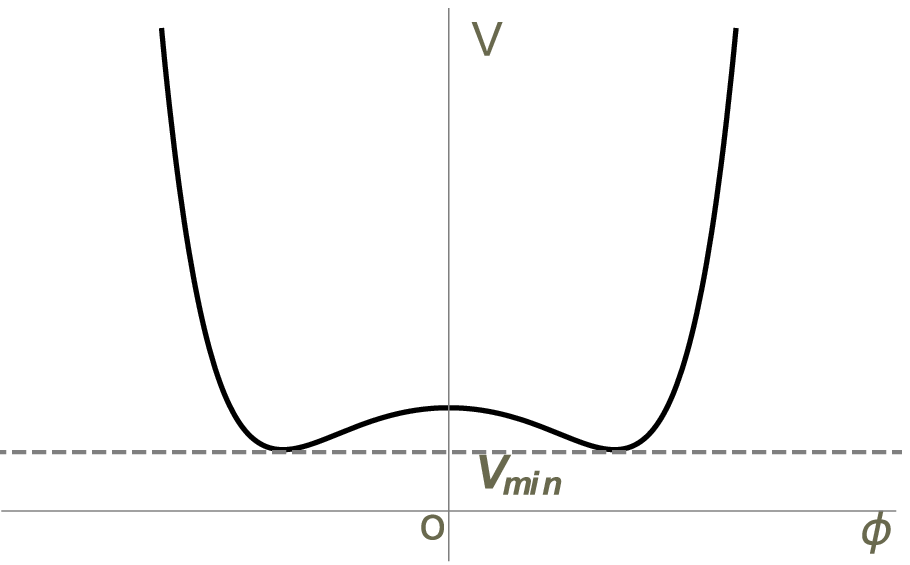}}\caption{Expansion
parameter, scale factor and scalar potential for $\omega=-1$, $\sigma=1$ and
$\alpha=0$.}%
\end{figure}\FloatBarrier

\section{Concluding remarks}

The recent discovery of the Higgs particle has clearly shown that scalar
fields play a fundamental role in the theory of subatomic physics. On the
other hand, as we have already remarked, a closer look at modern theoretical
cosmology reveals that scalar fields also have played an increasing important
role in the description of our universe. In particular, inflationary universes
and quintessence models for dark energy, respectively, resort to scalar fields
for explaining early expansion and cosmic acceleration. However, the nature of
the scalar field is still not known. It appears to us that a geometrical
scalar-tensor theory may provide a natural framework for investigating some
cosmological scenarios in which the scalar field is taken into account as an
essential ingredient for our description of the universe. With this
motivation, we have briefly examined some cosmological models generated by
different choices of the scalar field potential proposed in the literature. In
addition to obtaining some analytical solutions, we have constructed the phase
portrait of the solutions. In some cases we have found a cosmological scenario
which might be viewed as qualitatively describing a kind of geometric phase
transition of the universe.  The geometrical origin of the scalar field, which
is one of the basic tenets of the WIST's theoretical framework, appears as a
consequence of the application of the Palatini variational principle to the
gravitational sector of the action (\ref{action1}), the same powerful
principle that, when applied to Einstein-Hilbert action, leads directly to the
Riemannian nature of the space-time structure \cite{Ray}. The applications of
the Weyl geometrical scalar-tensor theories to cosmology naturally consider
cosmological scenarios where the presence of a scalar field is required. In
most models (inflation, dark energy, quintessence, etc) the real nature of the
scalar field is not known yet, and it is reasonably guessed that this kind of
phenomenological approach can be justified later. However, in the case of Weyl
geometrical scalar-tensor theory the nature of the scalar field is already
known from the beginning: it is part of the geometric framework of space-time.
In our view, this emphasis on the geometrical role of the scalar field seems
to be more in line with the program of geometrization of physics put forward
by Einstein when he conceived the general theory of relativity.

\section*{Acknowledgements}

\noindent The authors would like to thank CAPES and CNPq for financial
support. We are grateful to Dr. J. E. Madriz-Aguilar for reading the manuscript.


\begin{thebibliography}{99}                                                                                               %


\bibitem {Jordan}P. Jordan, \textit{Schwerkraft und Weltall} (Vieweg,
Braunschweig, 1955). For the history of scalar-tensor theories, see \ H.
Goenner, Gen. Rel. Grav. \textbf{44}, 2077 (2012). See also C. H. Brans, arXiv:gr-qc/0506063.

\bibitem {Brans}C. H. Brans and R. H. Dicke, Phys. Rev. \textbf{124}, 925
(1961). R. H. Dicke, Phys. Rev. \textbf{125}, 2163 (1962).

\bibitem {Faraoni}For a nice review of scalar-tensor theories, see Y. Fujii
and K. Maeda, \textit{The Scalar-Tensor Theory of Gravitation} (Cambridge
University Press, 2003). See also V. Faraoni, \textit{Cosmology in
Scalar-Tensor Gravity} (Kluwer Academic Publishers, 2004).

\bibitem {Peters}P. C. Peters, \textit{J. Math. Phys}. \textbf{10}, 1029
(1969). H. H. Soleng, \textit{Class. Quant. Grav.} \textbf{5}, 1489 (1988).
See also D. K. Sen, \textit{Z. Phys. }\textbf{149}, 311 (1957). G. Lyra, Math.
Z. \textbf{54}, 2. H. H. Soleng, \textit{Class. Quant. Grav.} \textbf{5},1489
(1988). See also D. K. Sen, \textit{Z. Phys. }\textbf{149}, 311 (1957). D. K.
Ross, Gen. Rel. Grav. \textbf{6}, 157 (1975).

\bibitem {Weyl}H. Weyl, Sitzungesber Deutsch. Akad. Wiss. Berlin, 465 (1918).
H. Weyl, \textit{Space, Time, Matter} (Dover, New York, 1952). (2006). See
also W. Pauli, \textit{Theory of Relativity} (Dover, New York, 1981). R.
Adler, M. Bazin and M. Schiffer, \textit{Introduction to General Relativity},
\ Ch. 15, (McGraw-Hill, 1975). L. O'Raiefeartaigh and N. Straumann, Rev. Mod.
Phys. \textbf{72}, 1 (2000).

\bibitem {Weyl geometry}For a more detailed account of Weyl geometry, see F.
Dahia, G.A.T. Gomez, C. Romero, J .Math .Phys. \textbf{49}, 102501 (2008). A
more formal treatment is given by G. B. Folland, J. Diff. Geom. \textbf{4},
145 (1970). For a clear and comprehensive review on Weyl geometry, see E.
Scholz, math.HO\ arXiv:1111.3220. See also E. Scholz, arXiv:1206.1559.

\bibitem {Novello}M. Novello, L.A.R. Oliveira, J.M. Salim, E. Elbas, Int. J.
Mod. Phys. \textbf{D1} (1993) 641-677.\ J. M. Salim and S. L. Saut\'{u},
Class. Quant. Grav. \textbf{13}, 353 (1996). H. P. de Oliveira, J. M. Salim
and S. L. Saut\'{u}, Class.Quant.Grav. \textbf{14},\textbf{\ }2833 (1997). V.
Melnikov, \textit{Classical Solutions in Multidimensional Cosmology} in
Proceedings of the VIII Brazilian School of Cosmology and Gravitation II
(1995), edited by M. Novello (Editions Fronti\`{e}res) pp. 542-560, ISBN
2-86332-192-7. K.A. Bronnikov, M.Yu. Konstantinov, V.N. Melnikov, Grav.Cosmol.
\textbf{1}, 60\textbf{\ }(1995). J. Miritzis, Class. Quantum .Grav.
\textbf{21}, 3043 (2004). J. Miritzis, J.Phys. Conf. Ser . \textbf{8},131
(2005). J.E.M. Aguilar and C. Romero, Found. Phys.\textbf{ 39} (2009)1205;
J.E.M. Aguilar and C. Romero, Int. J. Mod. Phys. A \textbf{24,} 1505 (2009).
\ C. Romero, J. B. Fonseca-Neto and M. L. Pucheu, Class.Quant.Grav.
\textbf{29, \ }155015 (2012). \ J. Miritzis, Int. J. Mod. Phys. D \textbf{22},
1350019 (2013). F. P. Poulis and J. M. Salim arXiv:1305.6830. R. Vazirian, M.
R. Tanhayi and Z. A. Motahar, Adv. High Energy Physics \textbf{7}, 902396
(2015). I. P. Lobo, A. B. Barreto, and C. Romero, Eur. Phys. J. C75
\textbf{9}, 448 (2015).

\bibitem {Pucheu}T. S. Almeida, M. L. Pucheu, C. Romero and J. B. Formiga,
Phys. Rev. D \textbf{89}, 064047.

\bibitem {Kamen}See for instance, A.Yu. Kamenshchik, E.O. Pozdeeva, A.
Tronconi, G. Venturi, S.Yu. Vernov, Class. Quantum Grav. \textbf{31}, 105003 (2014).

\bibitem {Matos}T. Matos, L. A. Urena-Lopez, Class. Quant. Grav. \textbf{17},
L75-L81 (2000). P. J. E. Peebles and A. Vilenkin, Phys. Rev. D \textbf{59}
063505 (1999).

\bibitem {Paliathanasis}See, for instance, A. Paliathanasis, M. Tsamparlis, S.
Basilakos, S. Capozziello, Phys. Rev. D \textbf{89}, 063532 (2014). A.
Paliathanasis, M. Tsamparlis, S. Basilakos, J. D. Barrow, Phys. Rev. D
\textbf{91}, 123535 (2015). A. Paliathanasis, M. Tsamparlis, S. Basilakos, J.
D. Barrow, Phys. Rev. D \textbf{93}, 043528 (2016).

\bibitem {Guth}A. H. Guth, Phys. Rev. D \textbf{23}, 347 (1981). See also, V.
Mukhanov, \textit{Physical Foundations of Cosmology} (Cambridge University
Press, Cambridge, 2005). A. R. Liddle and D. H. Lyth, \textit{Cosmological
Inflation and Large-Scale Structure}, (Cambridge University Press, Cambridge, 2000).

\bibitem {Linde}A. D. Linde, Phys. Lett. B\textbf{\ 108}, 389 (1982). A. D.
Linde, Phys. Lett. B\textbf{\ 129}, 177 (1983).

\bibitem {Planck}Planck Collaboration, Planck 2015 results. XX.
\textit{Constraints on inflation,} arXiv:1502.02114 [astro-ph.CO].

\bibitem {piran}T. Piran and R. M. Williams, Phys. Lett. B \textbf{163
}(1985). V. A. Belinsky, L. P. Grishchuck, I. M. Khalatnikov, and H. Sato,
\textit{Prog. Theor. Phys. }\textbf{79}, 676 (1988). V. A. Belinsky, L. P.
Grishchuck, I. M. Khalatnikov, and Ya. B. Zeldovich, \textit{Phys. Lett. B
}\textbf{155}, 232 (1985).

\bibitem {Stein}A. M. Levy, A. Ijjas, P. J. Steinhardt, Phys.Rev. D
\textbf{92} 6, 063524 (2015). For bouncing models in scalar-tensor theories
see also B. Boisseau, H. Giacomini D. Polarski and A.A. Starobinsky, JCAP 1507
(2015) 002.

\bibitem {Turok}See, for instance, P.J. Steinhardt and N. Turok, Phys. Rev. D
\textbf{65}, 126003 (2001). L. Baum and P.H. Frampton, Phys. Rev. Lett.
\textbf{98}, 071301(2007).

\bibitem {Asthekar}A. Ashtekar, Gen. Rel. Grav. \textbf{41}, 707 (2009).

\bibitem {Andronov}See, for instance, A. A. Andronov, E. A. Leontovich, I. I.
Gordon, and A. G. Maier, \textit{Qualitative theory of second-order dynamic
systems} (John Wiley \& Sons, New York, 1965).

\bibitem {Rubakov1}See, for a review, V. A. Rubakov, Phys. Usp. \textbf{57},
128 (2014), arXiv:1401.4024.

\bibitem {Arefeva1}I. Ya. Arefeva, L. V. Joukovskaya and S. Yu. Vernov, J.
Phys. A \textbf{41}, 304003 (2008), arXiv:0711.1364.

\bibitem {Arefeva2}I. Ya. Aref'eva, N. V. Bulatov and R. V. Gorbachev, Theor.
Math. Phys. \textbf{173}, 1466 (2012), arXiv: 1112.5951.

\bibitem {Lucchin}See, for instance, F. Lucchin and S. Matarrese, Phys. Rev. D
\textbf{32}, 1316 (1985).

\bibitem {Felder}G. Felder, A. Frolov, L. Kofman, and A. Linde, Phys. Rev. D
\textbf{66}, 023507 (2002). M. N. Verma, Astrophys. Space Sci. \textbf{161},
181 (1989).

\bibitem {Bazeia}D. Bazeia, C. B. Gomes, L. Losano and R. Menezes, Phys.Lett.
B \textbf{633}, 415 (2006).

\bibitem {Bezrukov}F. Bezrukov and M. Shaposhnikov, Phys. Lett. B
\textbf{659}, 703 (2008). I. Oda and T. Tomoyose, Adv. Stud. Theor. Phys.
\textbf{8}, 551(2014).

\bibitem {Arefeva3}I. Ya. Aref'eva, A. S. Koshelev and S. Yu. Vernov, Theor.
Math. Phys. \textbf{148}, 895 (2006), astro-ph/0412619. I. Ya. Aref'eva, A. S.
Koshelev and S. Yu. Vernov, Phys. Lett. B \textbf{628}, 1 (2005),
astro-ph/0505605. B. McInnes, Nucl. Phys. B \textbf{718}, 55, hep-th/0502209.

\bibitem {Ellis}See, for instance, G. F. R. Ellis and R. Marteens, Class.
Quantum Grav. \textbf{21}, 223 (2004). G.F.R. Ellis, J. Murugun and C.G.
Tsagas, Class. Quantum .Grav. \textbf{27}, 233 (2004).

\bibitem {Novello2}For a review, see M. Novello and S. Bergliaffa, Phys. Rep.
\textbf{463}, 127 (2008).

\bibitem {Biswas}T. Biswas, Mazumdar, W.Siegel, \ JCAP \textbf{603}, 9 (2006).
Yi-Fu Cai, D. A. Easson, R. Brandenberger, JCAP \textbf{1208}, 20 (2012).

\bibitem {Ray}See, for instance, J. R. Ray, Il Nuevo Cim. \textbf{25}, 706 (1975).
\end{thebibliography}
\end{document}